\documentclass[%
 reprint,
nofootinbib,
 amsmath,amssymb,amsfonts
 aps,
]{revtex4-1}

\usepackage{natbib}

\usepackage{xcolor}

\usepackage{bbold}
\usepackage{dsfont}
\usepackage{nccmath}

\usepackage[titletoc,toc,title]{appendix}
\usepackage{graphicx}% Include figure files
\usepackage{dcolumn}% Align table columns on decimal point
\usepackage{bm}% bold math
\usepackage[utf8]{inputenc}
\usepackage{amsmath}
\usepackage{amsfonts}
\usepackage{mathtools}
\usepackage{dsfont}
\usepackage{float} % Allows putting an [H] in \begin{figure} to specify the exact location of the figure
\usepackage{wrapfig} % Allows in-line images such as the example fish picture
\usepackage[caption=false]{subfig}

\usepackage{hyperref}

\usepackage{enumitem}

\usepackage{titlesec}

\titleformat{\paragraph}
{\normalfont\normalsize\bfseries}{\theparagraph}{1em}{}
\titlespacing*{\paragraph}
{0pt}{3.25ex plus 1ex minus .2ex}{1.5ex plus .2ex}

\begin{document}

%\preprint{APS/123-QED}

\title{Disformally Coupled Quintessence}% Force line breaks with \\
%\thanks{A footnote to the article title}%

\author{Elsa M. Teixeira$^{1,2}$, Ana Nunes$^3$, and Nelson J. Nunes$^1$}
\affiliation{%
$^1$Instituto de Astrof\'{i}sica e Ci\^{e}ncias do Espa\c{c}o, Faculdade de 
Ci\^{e}ncias da Universidade de Lisboa, Campo Grande, PT1749-016 Lisboa, Portugal\\
$^2$School of Mathematics and Statistics, University of Sheffield, Hounsfield Road, Sheffield S3 7RH, United Kingdom\\
$^3$BioISI - Biosystems and Integrative Sciences Institute, Faculdade de 
Ci\^{e}ncias da\\ Universidade de Lisboa, Campo Grande, PT1749-016 Lisboa, Portugal 
}%

\date{\today}% It is always \today, today,
             %  but any date may be explicitly specified
\begin{abstract}
In this work we consider a cosmological model in which dark energy is portrayed by a canonical scalar field which is allowed to couple to the other species by means of a disformal transformation of the metric. We revisit the current literature by assuming that the disformal function in the metric transformation can depend both on the scalar field itself and on its derivatives, encapsulating a wide variety of scalar-tensor theories. This generalisation also leads to new and richer phenomenology, explaining some of the features found in previously studied models. We present the background equations and perform a detailed dynamical analysis, from where new disformal fixed points emerge, that translate into novel cosmological features. These include early scaling regimes between the coupled species and broader stability parameter regions. However, viable cosmological models seem to have suppressed disformal late-time contributions. 
%\begin{description}
%\item[Usage]
%Secondary publications and information retrieval purposes.
%\item[PACS numbers]
%May be entered using the \verb+\pacs{#1}+ command.
%\item[Structure]
%You may use the \texttt{description} environment to structure your abstract;
%use the optional argument of the \verb+\item+ command to give the category of each item. 
%\end{description}
\end{abstract}

\pacs{Valid PACS appear here}% PACS, the Physics and Astronomy
                             % Classification Scheme.
%\keywords{Suggested keywords}%Use showkeys class option if keyword
                              %display desired
\maketitle
\onecolumngrid
%\tableofcontents

\section{\label{sec:int}Introduction}

Since 1998 we have been witnessing growing observational confirmation of the present accelerated expansion of the Universe \cite{acel1,acel2,Eisenstein:2005su,Aghanim:2018eyx}. This phenomenon can be attributed to an exotic fluid, the so-called Dark Energy (DE), which must amount to about 70$\%$ of the total content of the Universe, and whose effective negative pressure can successfully explain the observations (see \cite{Copeland:2006wr, Clifton:2011jh,  Bahamonde:2017ize} for recent reviews). Presently, the $\Lambda$CDM model is the most well-accepted cosmological model, consisting of a cosmological constant dark energy source, $\Lambda$, plus a dark matter component, needed in order to make formation of structure possible in the Universe \cite{cc1}. However, this paradigm faces some theoretical inconsistencies \cite{cc2}, motivating extensions of the concept of dark energy to a scalar field, with General Relativity as the underlying gravitational theory \cite{scalar1, scalar2,scalar3}. These scalar field based models, albeit simple, can give rise to very complex and rich phenomenologies, while making predictions that are testable according to observational constraints \cite{Aghanim:2018eyx}. In the plainest scenarios, DE is portrayed as a canonical scalar field, the \textit{quintessence} field, which does not interact with the other components in the Universe \cite{tsuq, Carroll:1998zi}. However, there is no fundamental reason to assume such a constraint and, in the simplest extension, the scalar field is allowed to couple non-minimally to the matter sector \cite{ELLIS1989264, Wetterich:1994bg, Amendola:1999qq, PhysRevD.62.043511,Nunes:2000ka,Gumjudpai:2005ry, Boehmer:2008av, Barros:2018efl, Barros:2019rdv}.

 One straightforward procedure for introducing a non-trivial coupling between the scalar field and matter is to consider that matter particles propagate in geodesics of a transformed metric, $\bar{g}_{\mu \nu}$, related to the gravitational metric, $g_{\mu \nu}$, by means of a field-dependent transformation. When this transformation corresponds to a rescaling of the metric we speak of conformal transformations, which affect the length of time-like and space-like intervals and the norm of time-like and space-like vectors while leaving the light cones unchanged:
 
 \begin{equation}
\bar{g}_{\mu \nu} =C (\phi) g_{\mu \nu},
\label{conf}
\end{equation}

\noindent where $C$ is the conformal factor. Conformal transformations are known to preserve the structure of Scalar-Tensor theories of the Brans-Dicke form \cite{Faraoni:1998qx}, which granted them an important role in contemporary gravitational theories. However, the conformal transformation is just the simplest way to relate two geometries. We could instead assume that the transformation depends not only on the scalar field itself but also on its first order partial derivatives:

\begin{equation}
 \bar{g}_{\mu \nu} =C ( \phi,X ) g_{\mu \nu} + D ( \phi,X ) \partial_{\mu} \phi \partial_{\nu} \phi,
\label{disf}
\end{equation}

\noindent where $C$ and $D$ are the conformal and disformal factors, respectively, which, in the most general case, depend on the field $\phi$ and on its corresponding kinetic term, $X=-\frac{1}{2} g^{\mu \nu} \partial_{\mu} \phi \partial_{\nu} \phi$. In this case, we speak of disformal transformations, which can not be interpreted as a metric rescaling, but rather as a stretching (or a compression) of the metric in a specific direction defined by the gradient of the scalar field, resulting in a distortion of both angles and lengths.

This formalism was first introduced by Bekenstein \cite{Bekenstein:1992pj}, while looking for the most general way to couple matter to the gravitational sector. 
But disformal transformations were only brought to the spotlight \cite{Zumalacarregui:2013pma} when it was shown that the form of the Horndeski Lagrangian is preserved under disformal transformations \cite{Bettoni:2013diz} with $C\equiv C ( \phi )$ and $D \equiv D (\phi )$ \cite{Zumalacarregui2013,Goulart:2013laa,Zumalacarregui:2013pma,Domenech:2015hka}. This has great physical importance, since Horndeski theories are the most general extensions of Scalar-Tensor theories of the Brans-Dicke type \cite{Horndeski:1974wa}. Its generalisation, known as Beyond Horndeski or GLPV theories, contain higher order derivative terms but are nonetheless healthy in the sense that they avoid instabilities \cite{Gleyzes:2014dya}. Analogously, it has been shown \cite{Gleyzes:2014qga} that the Lagrangian structure of GLPV models is preserved under disformal transformations of the form $C\equiv C  ( \phi )$ and $D \equiv D  ( \phi ,X )$, \cite{Langlois:2015cwa,Emond:2015efw,Crisostomi:2016czh}. However, if $C \equiv C( \phi ,X )$, terms that do not belong to the GLPV setting may arise, which are the cause of Ostrogatski instabilities \cite{Woodard:2015zca}. In \cite{Achour:2016rkg} it has also been shown that, under specific conditions, the form of the quadratic DHOST theories, is also preserved under disformal transformations of the general form presented in Eq.~\eqref{disf}. 

Disformal transformations in cosmology find fundamental motivation in brane-world models \cite{Koivisto:2013fta,
%,KOIVISTO:2013jwa,Koivisto:2014gia,Koivisto:2015vda,
Cembranos:2016jun} and massive gravity theories \cite{deRham:2010ik,deRham:2010kj} and have applications in several research fields such as varying speed of light models \cite{Clayton:1998hv,Clayton:1999zs},
%Bassett:2000wj,Magueijo:2003gj, 
relativistic MOND theories \cite{Bekenstein:2004ne,Skordis:2005xk},
%Skordis:2009bf,Milgrom:2009gv}, 
and extensions of dark matter \cite{Bettoni:2011fs,Deruelle:2014zza, Sebastiani:2016ras}
%Bettoni:2012xv,Deruelle:2014zza,Arroja:2015wpa}. 
They are also found in theories in which Lorentz invariance is broken spontaneously on a non-trivial background \cite{Brax:2012hm}, in Palatini formulations \cite{Olmo:2009xy} and have been used to investigate the onset of inflation in the early Universe \cite{Kaloper:2003yf,Germani:2011mx,vandeBruck:2015tna}.
% and the spherical collapse of overdense regions \cite{Sapa:2018jja}. 
%They have been been studied using PPN parameters \cite{Sakstein:2014aca,Ip:2015qsa,Ip:2015qsa} and in the language of differential forms \cite{Ezquiaga:2017ner}.
Disformal scalar field theories have been widely addressed in the context of dark energy scenarios \cite{Koivisto:2008ak,
%deRham:2010eu,Noller:2012sv
Zumalacarregui:2010wj,DeFelice:2011bh,Koivisto:2012za,Zumalacarregui:2012us,vandeBruck:2015ida,Bettoni:2015wla}, namely with dynamical systems techniques \cite{Sakstein:2014aca,Sakstein:2015jca,vandeBruck:2016jgg,Karwan:2016cnv}. Several laboratory experimental tests and cosmological observations have been proposed in order to constrain disformally coupled scalar field models, as for example in \cite{
%Brax:2012ie,vandeBruck:2013yxa,Brax:2013nsa,
Brax:2014vva,Brax:2015hma,vandeBruck:2015rma,Bettoni:2016mij,
%vandeBruck:2016hpz,vandeBruck:2017idm, 
Mifsud:2017fsy,Brax:2018bow,Dalang:2019fma}
%Leloup:2019fas} and references therein.
Disformal transformations in cosmological perturbations have been investigated, for example, in \cite{Minamitsuji:2014waa,
%Creminelli:2014wna,
Tsujikawa:2014uza,vandeBruck:2015ida}
%,Papadopoulos:2017xxx} 
 and in \cite{Motohashi:2015pra} it was shown that curvature perturbations are not identically invariant if $C \equiv C ( \phi ,X )$ in Eq.~\eqref{disf}.
Unlike conformal transformations, disformal transformations can change the causal structure of the spacetime and have non-trivial effects on radiation-like fluids, allowing for modifications in the behaviour of photons \cite{vandeBruck:2012vq,Brax2013}
%vandeBruck:2013yxa,Brax:2012ie,. 
%As this transformation is extremely general, these models can in principle be very interesting from the phenomenological point of view. Moreover, we have seen that most scalar tensor theories can be related to General Relativity with the addition of a disformally coupled matter sector and can be encapsulated in this formulation.
%Other phenomenology \cite{2012arXiv1201.2656D,Bettoni:2015wta,Koivisto:2015mwa,Minamitsuji:2016hkk,Vetsov:2018mld,BeltranJimenez:2018tfy,Firouzjahi:2018xob,Xiao:2018jyl,Domenech:2018vqj,Gannouji:2018aaw,Galtsov:2018xuc,Kuntz:2019zef,Geng:2019wgd,Hohmann:2019gmt,Delhom:2019yeo}.????

The main goal of this work is to perform a detailed analysis of generalised couplings between a canonical scalar field, portraying dark energy, and the matter sector. To do so, we assume that the scalar field is disformally coupled to a perfect fluid, as was done in \cite{nelson}, but we extend the analysis in the existing literature by assuming that the disformal coefficient can also depend on the kinetic term associated to the scalar field. This problem has already been addressed in \cite{tai} but we intend to revisit it with a different approach. We assume that different matter species couple differently to the scalar field as couplings to baryons and photons are severely constrained from Solar System tests \cite{Will:2014kxa,Sakstein:2014isa,Ip:2015qsa}. This choice of non-universal coupling is motivated by the recent constraints on the speed of gravitational waves inferred from observations of the optical counterpart of a binary neutron star merger \cite{Monitor:2017mdv}.

This paper is organised as follows. In Sec. II, we introduce the kinetic disformally coupled quintessence scenario and derive the general equations of motion. In Sec. III we present the equations in a Friedmann-Lema\^{i}tre-Robertson-Walker (FLRW) background and the form of the interaction term. In Sec. IV we follow to rewrite the equations as a dynamical system in terms of dimensionless variables with physical interest and proceed to the study of the single-fluid scenario for both pressureless and relativist fluids. Additionally we comment on the novelties associated with the introduction of the kinetic dependence, in contrast with \cite{nelson} and also comment on some of the ``pathologies" and features found in previous works. In Sec. V we discuss the existence of viable scenarios with cosmological interest. We perform numerical simulations that suggest that while the emergence of the disformal fixed points can lead to early scaling regimes, the disformal contribution is always suppressed at late times. Finally, we conclude in Sec. VI.

\section{The Model} \label{sec:disfmod}

In this section we will focus on the construction of a cosmological framework in the presence of a dark energy component, that is allowed to interact with other matter species. This coupling emerges naturally in the theory by means of a disformal transformation of the metric tensor from the Einstein frame to the Jordan frame:

\begin{equation}
g_{\mu \nu} \longmapsto \bar{g}_{\mu \nu} =C ( \phi ) g_{\mu \nu} + D (\phi,X) \partial_{\mu} \phi \partial_{\nu} \phi,
\label{transdd}
\end{equation}

\noindent in which case, we can also derive the inverse transformed metric

\begin{equation}
g^{\mu \nu} \longmapsto \bar{g}^{\mu \nu} = \frac{1}{C} g^{\mu \nu} -\frac{D \partial^{\mu} \phi \partial^{\nu} \phi}{C^2-2 C D X} ,
\label{inversedd}
\end{equation}

\noindent and the determinant of the metric

\begin{equation}
g \longmapsto \bar{g}= C^{3} \left( C-2 D X \right) g,
\end{equation}

\noindent where $C \equiv C ( \phi )$ and $D \equiv D ( \phi,X )$ are the conformal and disformal coupling functions, respectively, and $X=-\frac{1}{2} g^{\mu \nu} \partial_{\mu} \phi \partial_{\nu} \phi$ is the kinetic term associated with the scalar field. 
%The cosmological implications of the interaction between dark energy and matter by virtue of disformal transformations have been vastly studied for the case where both the conformal and disformal coefficients depend only on the field $\phi$ but generically, the functions $C$ and $D$ can also depend on the kinetic term $X$. 
%, through a four-dimensional spacetime manifold $\mathcal{M}$ endowed with a metric $g_{\mu \nu}$,

Hereafter we identify the field $\phi$ in Eq. \eqref{transdd} as the dark energy field and, therefore, the coupling between matter and dark energy is fully described by considering that each matter fluid propagates on geodesics defined according to some metric disformally related to the gravitational metric. Thereupon, we consider an Einstein frame action enclosing the gravitational Lagrangian and the scalar field Lagrangian, which depend on the metric $g_{\mu \nu}$, and the matter Lagrangian, a function of the disformal metric $\bar{g}_{\mu \nu}$, as defined in Eq.~\eqref{transdd}:

\begin{equation}
S =\int d^4x \sqrt{-g} \left[\frac{1}{2 \kappa^2} R + X - V ( \phi ) \right] 
+ \int d^4x \sqrt{-\bar{g}} \bar{\mathcal{L}} \mathopen{} \left( \bar{g}_{\mu \nu}, \psi, \partial_{\mu} \psi \right) \mathclose{},
\label{actiond}
\end{equation}

\noindent where the first term is the usual Einstein-Hilbert form for the gravitational action, constructed from the scalar curvature, $R$, with $\kappa^2 \equiv 8 \pi G$ being the scaled gravitational constant. The term $\mathcal{L}_{\phi} \equiv X - V ( \phi )$ is the Lagrangian density of the canonical dark energy field, where $V(\phi )$ is a general self-coupling potential. $\bar{\mathcal{L}}$ stands for the Lagrangian of the matter fluids, where $\psi$ denotes matter fields propagating on geodesics defined by $\bar{g}_{\mu \nu}$. As a first approximation, and for the sake of simplicity of the analysis, we focus on couplings of the dark energy field to a single effective fluid. 

The Einstein field equations are derived by variation of the action in Eq.~\eqref{actiond} with respect to the gravitational metric $g_{\mu \nu}$:

\begin{equation}
G_{\mu \nu} \equiv R_{\mu \nu} - \frac{1}{2} g_{\mu \nu} R = \kappa^2 \mathopen{}\left(T_{\mu \nu}^{\phi} + T_{\mu \nu}  \right) \mathclose{},
\label{efeqd}
\end{equation}

\noindent where $G_{\mu \nu}$ is the Einstein tensor and $R_{\mu \nu}$ is the Ricci curvature tensor. The objects $T_{\mu \nu}^{\phi}$ and $T_{\mu \nu}$ stand for the energy-momentum tensors for the scalar field $\phi$ and matter:

\begin{equation}
T_{\mu \nu}^{\phi} \equiv -\frac{2}{\sqrt{-g}} \frac{\delta \left( \sqrt{-g} \mathcal{L}_{\phi} \right)}{\delta g^{\mu \nu}}, \ \ \  T_{\mu \nu} \equiv- \frac{2}{\sqrt{- g}} \frac{\delta \left( \sqrt{- \bar{g}} \bar{\mathcal{L}} \right)}{\delta g^{\mu \nu}} .
\label{Tcd}
\end{equation} 
The energy-momentum tensor for quintessence can be computed directly from Eq.~\eqref{actiond}:

\begin{equation}
T_{\mu \nu}^{\phi}= \partial_{\mu} \phi \partial_{\nu} \phi - g_{\mu \nu} \mathopen{} \left( \frac{1}{2} g_{\alpha \beta} \partial^{\alpha} \phi \partial^{\beta} \phi + V (\phi ) \right) \mathclose{}.
\label{tmunudd}
\end{equation}
 
The energy momentum tensor in the transformed frame is related to the one in the original frame defined in Eq.~\eqref{Tcd} as

\begin{align}
T^{\alpha \beta} = \frac{\sqrt{-\bar{g}}}{\sqrt{-g}} \frac{\delta\bar{g}_{\rho \delta}}{\delta g_{\alpha \beta}} \bar{T}^{\rho \delta} = \left(C \delta^{\alpha}_{\rho} \delta^{\beta}_{\delta} - \frac{1}{2} {D}_{,X} \partial^{\alpha} \phi \partial^{\beta} \phi \partial_{\rho} \phi \partial_{\delta} \phi \right) J  \bar{T}^{\rho \delta}, 
\label{tem}
\end{align}

\noindent where $J\equiv \sqrt{-\bar{g}}/\sqrt{-g}$ is the Jacobian of the transformation, the subscript ($,$) denotes partial derivatives and, accordingly

\begin{equation}
\bar{T}_{\mu \nu} \equiv- \frac{2}{\sqrt{- \bar{g}}} \frac{\delta \left( \sqrt{- \bar{g}} \bar{\mathcal{L}} \right) }{\delta \bar{g}^{\mu \nu}} .
\end{equation}

The equation of motion, or the \textit{coupled Klein-Gordon equation}, is derived through variation of the action with respect to $\phi$ and reads:

\begin{equation}
\Box \phi = V_{, \phi} - Q,
\label{eqmovcan}
\end{equation}

\noindent where $\Box=g^{\mu \nu}\nabla_{\mu} \nabla_{\nu}$ is the D'Alembertian operator and

\begin{equation}
Q = \frac{1}{2} \nabla_{\sigma} \left( J  \partial^{\sigma} \phi \bar{T}^{\mu \nu} {D}_{,X} \partial_{\mu} \phi \partial_{\nu} \phi \right) + \frac{1}{2} J \bar{T}^{\mu \nu} \left({C}_{, \phi} g_{\mu \nu} + {D}_{, \phi} \partial_{\mu} \phi \partial_{\nu} \phi\right) - \nabla_{\mu} \mathopen{} \left(J \bar{T}^{\mu \nu} D \partial_{\nu} \phi \right)\mathclose{}, \label{Qd}
\end{equation}

\noindent is the interaction term, responsible for establishing how energy flows between the dark energy component and the matter sector. From Eqs.~\eqref{tmunudd} and~\eqref{eqmovcan}, we derive the following conservation relations, needed in order to preserve general covariance:

\begin{equation}
\nabla_{\mu} T^{\mu \nu}= Q \partial^{\nu} \phi,
\label{qnabd}
\end{equation}
\noindent and 
\begin{equation}
\nabla_{\mu} T^{\mu \nu}_{\phi}= - Q \partial^{\nu} \phi.
\label{qnabf}
\end{equation}
Note that the Einstein tensor $G_{\mu \nu}$ is indeed divergenceless but that does not imply the individual conservation of the energy-momentum tensors. This is ascribed to the fact that the term $\bar{\mathcal{L}}$ depends on the field $\phi$, through $\bar{g}_{\mu \nu}$.

For the trace of the fluid's energy momentum tensor we have:

\begin{equation}
T=g_{\alpha \beta} T^{\alpha \beta}= J C g_{\rho \delta} \bar{T}^{\rho \delta}+ J {D}_{,X} X \partial_{\rho} \phi \partial_{\delta} \phi \bar{T}^{\rho \delta}.
\label{tctransd}
\end{equation}

Using Eq.~\eqref{tctransd}, it is possible to rewrite Eq.~\eqref{tem}, from where we extract the following relation for the energy momentum tensor in the transformed frame:

\begin{equation}
 \bar{T}^{\alpha \beta}= \frac{T^{\alpha \beta}}{C J}+\frac{\partial^{\alpha} \phi \partial^{\beta} \phi {D}_{,X}  }{2 C J \left( C-2 {D}_{,X} X^2 \right)} \partial_{\rho} \phi \partial_{\delta} \phi T^{\rho \delta}.
 \label{talphd}
\end{equation}

Using the relations in Eqs.~\eqref{tctransd} and~\eqref{talphd} it is possible to write Eq.~\eqref{Qd} in terms of quantities in the unbarred frame.

\section{Background Cosmology} \label{sec:backd}

Having presented the general formalism of the kinetic disformally coupled model, we next specify a background cosmological framework. We assume a homogeneous, isotropic, spatially flat FLRW metric in Cartesian coordinates:
\begin{equation}
ds^2= g_{\mu \nu} dx^{\mu} dx^{\nu} =- dt^2+a^2 (t )\delta_{ij} dx^i dx^j,
\label{frwmetricc}
\end{equation}
\noindent where $a (t)$ is the scale factor, a function of the cosmic time $t$.
The homogeneous scalar field is assumed to depend on time only, and time derivatives in this frame will be denoted by an upper dot. 

We also assume that the coupled species under consideration is well described, at large scales and with high precision, by a continuous perfect fluid, so that

\begin{equation}
T_{\mu \nu} = \left( \rho + p \right) u_{\mu} u_{\nu} + p g_{\mu \nu},
\label{tperfd}
\end{equation}

\noindent where $\rho$ and $p$ are the fluid's energy density and pressure, respectively, and $u_{\mu}$ is the fluid's four-velocity, which for a comoving observer is given by $u_{\mu}= \left( -1,0,0,0 \right)$.

The energy momentum tensor of the field can also be rewritten so as to convey a perfect fluid form:

\begin{equation}
T_{\mu \nu}^{\phi} = \left( \rho_{\phi} + p_{\phi} \right) u_{\mu} u_{\nu} + p_{\phi} g_{\mu \nu},
\label{tfperfd}
\end{equation}

\noindent as long as, following Eq.~\eqref{tmunudd}, the field's energy density and pressure are expressed as:

\begin{equation}
\rho_{\phi}= \frac{1}{2} \dot{\phi}^2 + V (\phi)\ \ \ \text{and}\ \ \ p_{\phi}=\frac{1}{2} \dot{\phi}^2 - V ( \phi).
\label{rpphiq}
\end{equation}
Also, according to Eq.~\eqref{rpphiq}, we derive the equation of state (EoS) parameter for the field:

\begin{equation}
w_{\phi}=\frac{p_{\phi}}{\rho_{\phi}} = \frac{\frac{1}{2} \dot{\phi}^2 - V ( \phi )}{\frac{1}{2} \dot{\phi}^2 + V( \phi)}.
\label{esfield}
\end{equation}
%
% From Eq.~\eqref{esfield} it is straightforward to conclude that, in the limit where the potential energy prevails, we approach a cosmological constant-type scenario, with $w_{\phi}=-1$, and, on the other hand, when the evolution is governed by the kinetic energy, the EoS becomes $w_{\phi}=1$. Hence, the value of $w_{\phi}$ varies as a function of the kinetic energy, $\frac{1}{2} \dot{\phi}^2$, and the potential energy, $V \left( \phi \right)$: $-1 \leq w_{\phi} \leq 1$.
 
The Einstein field equations give rise to two coupled differential equations for the scale factor $a (t)$ and the functions $\rho( t )$ and $p( t)$. Additionally, taking Eq.~\eqref{eqmovcan} and Eq.~\eqref{efeqd}, we may write the modified Klein-Gordon equation, the fluid conservation equation and the Friedmann equations for the single-fluid model in this frame:

\begin{equation}
\ddot{\phi} + 3H \dot{\phi}+V_{, \phi}=Q,
\label{kgd}
\end{equation}

\begin{equation}
\dot{\rho} + 3 H \rho \left( 1+ w \right) = -Q \dot{\phi} ,
\label{fluidcond}
\end{equation}

\begin{equation}
H^2= \frac{\kappa^2}{3} \mathopen{} \left( \rho_{\phi}+ \rho \right) \mathclose{},
\label{fried1d} 
\end{equation}

\begin{equation}
\dot{H}=- \frac{\kappa^2}{2} \mathopen{} \left[ \rho_{\phi} \left( 1+w_{\phi} \right) + \rho \left(1+ w \right) \right] \mathclose{},
\label{fried2d}
\end{equation}

\noindent where $H \equiv \dot{a}/a$ is the Hubble rate, $w= p/\rho$ is the EoS parameter for the coupled fluid and $Q$ is the interaction term in this frame, as defined in Eq.~\eqref{Qd}. 
The exchange of energy becomes more apparent when writing the continuity equation for the field, derived from Eqs.~\eqref{qnabf} and~\eqref{kgd}:

\begin{equation}
\dot{\rho_{\phi}} + 3 H \rho_{\phi} \left( 1+ w_{\phi} \right) =  Q \dot{\phi}.
\label{fieldcond}
\end{equation}

\noindent Together, Eqs.~\eqref{fieldcond} and~\eqref{fluidcond} define the direction in which energy is being transferred: if $Q \dot{\phi} >0$, it is the matter sector which grants energy to the DE field, whereas if $Q \dot{\phi}<0$ it is the dark energy fluid which sources the matter sector.

Ultimately, we find that the interaction term for an FLRW cosmology with a perfect fluid coupled to the scalar field, reads:

\begin{equation}
Q= \frac{\mathcal{A}}{\mathcal{B}},
\label{qi}
\end{equation}

\noindent where

\begin{gather}
\mathcal{A} = \frac{1}{2} \frac{C_{, \phi}}{C} \left(3 p -\rho \right)+  \frac{D}{C} \biggl\lbrace  V_{, \phi}  \rho + 3 H \left( \rho + p \right) \dot{\phi} + \frac{C_{, \phi}}{C} \left[ 2  \rho +3 \frac{D_{,X} X}{D}  \left(\rho-2 p \right) \right] X - \frac{D_{, \phi}}{D} \rho X \nonumber \\
-\, 2  \frac{D_{, X \phi} X}{D}  \rho X +  \frac{D_{,X} X}{D} \left( 3H \left( 5  \rho +p \right) \dot{\phi} + 5  \rho V_{, \phi} \right) +  \frac{D_{,XX} X^2}{D}  \left(6H \rho \dot{\phi} +2 \rho V_{, \phi} \right) \biggr\rbrace \nonumber \\
+ \left( \frac{D}{C}\right)^2 \biggl\lbrace  \frac{D_{,X} X}{D}  \left[6  V_{, \phi}  \rho  X + 6H \left(3 \rho - p \right) X  \dot{\phi} + 2  \frac{D_{, \phi}}{D} \rho X^2  + \frac{D_{,X} X}{D}  \left(-6H \left( \rho +p \right) X  \dot{\phi} \right. \right. \nonumber \\
 \left. \left. +\,  6 \frac{C_{, \phi}}{C}  p  X^2  - 2 V_{, \phi} \rho  X \right) \right] - 4 \frac{D_{,X \phi} X}{D}  \rho X^2 +  \frac{D_{,XX} X^2}{D}  \left(12 H \rho X \dot{\phi} + 4  V_{, \phi} \rho  X \right) \biggl\rbrace, \label{bi}
\end{gather}

\noindent and 

\begin{gather}
\mathcal{B} =\, 1+ \frac{D}{C}\bigg[ \rho- 2X + \frac{{D}_{,X} X}{D} \left( 5 \rho -6X \right)+2 \frac{ {D}_{,XX} X^2}{D}  \rho \bigg] +\left( \frac{D}{C} \right)^2 \bigg[  \frac{{D}_{,X} X}{D} \left(6 \rho + 4X \right) X  \nonumber \\
+  \left(\frac{{D}_{,X} X}{D} \right)^2 \left( 8X-2 \rho \right) X + 4  \frac{{D}_{,XX} X^2}{D} \rho X \bigg]. \label{ai}
\end{gather}

\section{Dynamical System} \label{sec:disfdyneq}

The system of equations~\eqref{kgd}-\eqref{fried2d} can be expressed as a set of first order differential equations. When doing so, it is useful to introduce the following dimensionless variables \cite{Copeland:1997et}:

\begin{gather}
 x^2\equiv \frac{\kappa^2 \phi'^2}{6},\ \ \ y^2\equiv\frac{\kappa^2 V}{3 H^2},\ \ \ z^2\equiv \frac{\kappa^2 \rho}{3 H^2},\ \ \ \sigma\equiv\frac{D H^2}{C \kappa^2 },\ \ \ \lambda_V\equiv - \frac{1}{\kappa} \frac{V_{, \phi}}{V},\ \ \ \lambda_C \equiv - \frac{1}{\kappa} \frac{{C}_{, \phi}}{C}, \nonumber \\
\lambda_D \equiv - \frac{1}{\kappa} \frac{{D}_{, \phi}}{D},\ \ \ \mu_D \equiv \frac{{D}_{, X}X}{D},\ \ \ \eta_D \equiv \frac{{D}_{,XX}X^2}{D} ,\ \ \  \xi_D \equiv- \frac{1}{\kappa} \frac{{D}_{, \phi X}X}{D},
\label{variablesd}
\end{gather}

\noindent where the Einstein frame time coordinate $t$ has been replaced by the number of e-folds, $N \equiv \ln a$, and derivatives with respect to $N$ are denoted by a prime.

For concreteness, we assume that the conformal and disformal coupling functions and the scalar field potential have the following forms:

\begin{equation}
C ( \phi)= C_0 e^{2 \alpha \kappa \phi},\ \ D ( \phi,X )=\frac{e^{2 \left( \alpha+\beta \right) \kappa \phi}}{M^{4+4\mu}} X^{\mu}\ \  V ( \phi )= V_0 e^{- \lambda \kappa \phi},
\label{cdv}
\end{equation}

\noindent where $\alpha$, $\beta$, $\lambda$ and $\mu$ are all dimensionless constant parameters and $M$, $V_0$ and $C_0$ are constant parameters with dimensions of mass, mass to the fourth, and no dimensions, respectively. From a physical point of view, this choice is interesting because exponential forms for the potential and the coupling function are known to give rise to the simplest non-trivial closed set of dynamical equations, which lead to the emergence of late time accelerated solutions that describe viable dark energy scenarios. In this context, the quantities $\lambda_V$ to $\xi_D$, defined in~\eqref{variablesd}, become constants. 
Since the parameter $C_0$ is merely a global rescaling of the conformal contribution, without loss of generality, we may set $C_0=1$. The variable $\sigma$ defines the disformal strength and can be expressed in terms of the scale of the disformal function, $M$ as:

\begin{equation}
\sigma = \frac{H^2}{\kappa^2} \frac{e^{2 \beta \kappa \phi} X^{\mu}}{M^{4+4\mu}}.
\end{equation}

It is noteworthy to highlight that the novelty associated with this model lies in the possibility of having the disformal function depend on the kinetic term of the scalar field. We restrict the model to the case of a power-law dependence, expressed through the new parameter $\mu$ and moreover, we will consider only positive values of $\mu$ in order to avoid possible singularities in the coupling (when the velocity of the scalar field, \textit{i.e.} $X$, goes through zero). We are interested in understanding to what extent the addition of this degree of freedom affects the overall dynamics of the system.
Comparing the definitions in~\eqref{variablesd} and~\eqref{cdv}, it is easy to conclude that the freedom associated to the system lies in 4 parameters only:

\begin{equation}
\lambda_V=\lambda,\ \ \ \lambda_C= -2 \alpha,\ \ \ \lambda_D=-2 \left( \alpha +\beta \right),\ \ \ \mu_D=\mu,\ \ \ \eta_D= \mu \left( \mu-1 \right),\ \ \ \xi_D=-2\mu \mathopen{} \left( \alpha +\beta \right) \mathclose{}.
\end{equation}

From the expression for $Q$ given in Eq.~\eqref{qi} (for a general form of the potential and the coupling functions), the definition of the dynamical variables in~\eqref{variablesd} and~\eqref{cdv}, the interaction term can ultimately be rewritten, as a function of the parameters, $\alpha$, $\beta$, $\lambda$ and $\mu$, as:

\begin{gather}
\left\lbrace 1+ 3 \sigma \left[ \left(1+3\mu+2\mu^2 \right) z^2 - 2x^2 \left(1+3 \mu \right) \right] + 18 \sigma^2 \left[ \left(1+\mu \right) \mu z^2 +2 x^2 \mu \left(1+2\mu \right) \right] x^2 \right\rbrace Q = \frac{3 H^2 z^2}{\kappa} \bigg\lbrace - \alpha \left(1-3w \right) \nonumber \\
  + 3 \sigma \left[ \sqrt{6} x \left(1+3 \mu+ 2\mu^2 + w \left( 1+ \mu \right) \right) + x^2 \left( 4 \alpha + 6 \alpha \mu \left(1-2 w \right) - 2 \left( \alpha + \beta \right) \left(1+2 \mu \right) \right) - y^2 \lambda \left( 1+ 3 \mu +2\mu^2 \right) \right]  \nonumber \\
 + 18 \sigma^2 \left[ \sqrt{6} x^3 \left( \mu + \mu^2 - w \mu \left(1+\mu \right) \right) + x^4 \left(-2 \mu \left( \alpha + \beta \right) +6 w \alpha  \mu^2 \right) - x^2 y^2 \lambda \mu \left( 1 + \mu \right) \right] \bigg\rbrace.
\label{qpar}
\end{gather}

\noindent As expected, by setting $\mu=0$ in Eq.~\eqref{qi} we recover the coupling function introduced in \cite{nelson}, for a purely field-dependent disformal transformation. One immediate conclusion is that the coupling function given in Eq.~\eqref{qpar} has a second order dependency on the variable related to the disformal coupling, $\sigma$, whereas in \cite{nelson} only a linear dependence was found. As we will see, this results in the emergence of new (disformal) fixed points solutions.

With this parametrisation, the system is promptly related to previously studied cases:

\begin{itemize}
\item If $\mu=0$ the system reduces to disformally coupled quintessence, in the case where the conformal and disformal functions depend only on the field itself. This was studied in \cite{nelson, vandeBruck:2015ida}.

\item If $M^{-1} = 0$ this system coincides with the standard coupled quintessence scenario, which was properly analysed in \cite{Holden:1999hm,PhysRevD.62.043511, Gumjudpai:2005ry, Bahamonde:2017ize}.

\item Finally, if $M^{-1} = 0$ and $\alpha=0$ we are left with the standard (uncoupled) quintessence scenario, presented in \cite{Copeland:1997et, Bahamonde:2017ize}.

\end{itemize}

Taking the variables defined in~\eqref{variablesd}, along with the choice for the form of the coupling functions and the potential in~\eqref{cdv}, we can write the system of dynamical equations for the single fluid case, which is guaranteed to be autonomous and closed:

\begin{eqnarray}
x'&=&-\left( 3 + \frac{H'}{H}\right) x + \sqrt{\frac{3}{2}} \mathopen{} \left( \lambda y^2 + \frac{\kappa Q}{3 H^2}\right) \mathclose{},
\label{xld} \\
y'&=&-\sqrt{\frac{3}{2}} \left(\lambda x+ \sqrt{\frac{2}{3}} \frac{H'}{H}\right) y,
\label{yld} \\ 
z'&=&-\frac{3}{2} \left(1 + w + \frac{2}{3} \frac{H'}{H} + \frac{1}{3} \sqrt{\frac{2}{3}} \frac{\kappa Q}{H^2} \frac{x}{z^2} \right) z,
\label{zld} \\
\sigma ' &=&2 \left[\sqrt{6} \beta x +  \mu \frac{x'}{x} +  \frac{H'}{H} (1+ \mu) \right] \sigma ,
\label{sild}
\end{eqnarray}

\noindent where 

\begin{equation}
\frac{H'}{H}= - \frac{3}{2} \left[ 2x^2 + (1+w)z^2 \right] = - \frac{3}{2} \mathopen{} \left( 1+w_{\text{eff}} \right) \mathclose{}.
\label{hlinhad}
\end{equation}

\noindent Equations \eqref{xld}-\eqref{hlinhad} can be simplified according to the Friedmann constraint,

\begin{equation}
x^2 + y^2 + z^2=1,
\label{fc}
\end{equation}

\noindent which we use to write $z$ in terms of the other variables, reducing the dimensionality of the system. 
The quantity in brackets in Eq.~\eqref{hlinhad} defines the effective equation of state parameter:

\begin{equation}
 w_{\text{eff}}= x^2 - y^2 + \left(1-x^2-y^2 \right) w.
\label{weffd}
\end{equation}

Recall that the effective EoS parameter is the one from which we gather if the Universe portrays a period of accelerated ($w_{\text{eff}} < -1/3$) or decelerated ($w_{\text{eff}} > -1/3$) expansion at present times. 

Time integration of Eq.~\eqref{hlinhad} gives the evolution of the scale factor over time, at any fixed point $\left( x^f,y^f,\sigma^f \right)$ of the phase space:

\begin{equation}
a \propto t^{\frac{2}{3 \left( 1 + w_{\rm eff}^f \right)}},
\label{adisf}
\end{equation}

\noindent with $w_{\rm eff}^f  \equiv w_{\rm eff} \left( x_f, y_f, \sigma_f \right)$ as defined in Eq.~\eqref{weffd}. Note that this means that even if we are not able to describe the entire evolution of the Universe, the asymptotic behaviour will always be well-defined, provided that it is given by a specific fixed point solution.

It is also possible to express the relative energy density and the equation of state parameter of the field, in terms of the dynamical variables defined in \eqref{variablesd}:

\begin{equation}
\Omega_{\phi}= x^2 + y^2,
\label{omegad}
\end{equation}

\begin{equation}
w_{\phi}= \frac{x^2 - y^2}{x^2+y^2}.
\label{wd}
\end{equation}

\noindent The EoS parameter of the fluid in the Jordan frame, defined by the metric $ \bar{g}_{\mu \nu}$, reads:

\begin{equation}
w= \bar{w} \frac{1-6 \sigma x^2}{1-6 \mu \sigma x^2}.
\label{wtransd}
\end{equation}

\noindent Note that $\bar{w}$ is frame-dependent, except for the exceptional case of the system with $\mu=1$. For example, in the case of pressureless matter and radiation, $ \bar{w}_{m}=0$ and $\bar{w}_{r} = 1/3$ respectively. It follows that for dust-like (pressureless) fluids, the EoS vanishes in both frames, whereas for radiation-like fluids in the Einstein frame it becomes:

\begin{equation}
w_{r}= \frac{1}{3}\frac{1-6 \sigma x^2}{1-6 \mu \sigma x^2}.
\label{wtransdr}
\end{equation}

We also perform the following parametrisation:
 
\begin{equation}
 \gamma \equiv \bar{w} +1,
\end{equation}
 
\noindent such that $0 \leq \gamma \leq 2$.

%From Eq.~\eqref{fc}, the meaning of the dynamical variables $x$ and $y$ is straightforward: $x^2$ plays the role of the relative kinetic energy density of the field, while $y^2$ stands for the potential energy density. Together they compose the total energy density parameter of the scalar field, given by Eq.~\eqref{omegad}.

\subsection{Phase Space and Invariant Sets}

The physical phase space can be restricted by considering that the energy density of the matter fluids is always positive, implying that $0\leq \Omega_{\phi} \leq 1$, which translates into:

\begin{equation}
0 \leq x^2+y^2 \leq 1.
\end{equation}

\noindent This condition defines a unitary circle on the $\left( x,y \right)$-plane, centred at the origin.
According to Eq.~\eqref{omegad}, points on the unit circle stand for scalar field dominated configurations (\textit{i.e.}, points for which $\Omega_{\phi} = 1$). 

Examining the dynamical system of equations~\eqref{xld},~\eqref{yld} and~\eqref{sild}, we conclude that $y=0$ is an invariant set of the system. Furthermore, the system is invariant under the transformation $y \longmapsto -y$.
For this reason, throughout this work, we focus only on positive values of $y$, reducing the physical phase space in the $\left(x,y\right)$-plane to a half-unit disk centred at the origin. 

The $\left( x,y,\sigma \right)$-phase space is still non-compact since:

\begin{equation}
0\leq \sigma < +\infty.
\end{equation}

\noindent This means that, in order to draw the full space, we first need to compactify it.To do so, we introduce the variable (following the procedure in \cite{nelson})

\begin{equation}
\Sigma \equiv \arctan \sigma.
\label{compatd}
\end{equation} 

\noindent The phase space is now the compact set

\begin{equation}
-1\leq x \leq1,\ \  0\leq y \leq\sqrt{1-x^2},\ \ 0\leq \Sigma < \pi /2,
\label{exisd}
\end{equation}

\noindent a semi-circular prism of length $\pi/2$. This parametrization is useful in order to draw the global phase space, including the asymptotic behaviour as $\sigma \rightarrow \infty$.

In order to assure that the disformal transformation is well defined (invertible and real), we take into account the parameter $Z$ (see Appendix A of \cite{Zumalacarregui:2012us}), related to the Jacobian $J$, in general, as

\begin{equation}
J=\frac{\sqrt{-\bar{g}}}{\sqrt{-g}} = C^2 \sqrt{1 + \frac{D}{C} g^{\mu \nu} \partial_{\mu} \phi \partial_{\nu} \phi}= C^2 Z,
\label{jota}
\end{equation}

\noindent which, in terms of the variables defined in~\eqref{variablesd}, reads

\begin{equation}
Z=\sqrt{1-6 \sigma x^2}.
\label{zd}
\end{equation}

\noindent From Eq.~\eqref{jota}, we require $Z \in \mathds{R} \setminus \lbrace 0 \rbrace$. Even though it may be problematic, we do not fully disregard the $Z=0$ case, which stands for a singularity in the metric. On these grounds, the phase space is further restricted by:

\begin{equation}
6 \sigma x^2 \leq 1.
\label{exiss}
\end{equation}

\noindent This condition implies that the compactified surface $\Sigma \rightarrow \pi/2$ (equivalent to $\sigma \rightarrow \infty$) intersects the phase space on the line $x = 0$ only.

We still have to account for the existence of an additional singularity, related to the limit in which the denominator of Eq.~\eqref{qi} becomes null and, consequently, Eq.~\eqref{xld} diverges. This corresponds to a surface which, by appropriate numerical simulations, we find to always lie outside the phase space delineated by~\eqref{exiss}, meaning that the coupling $Q$ never becomes singular in this regime.

The system is invariant under the simultaneous transformation $\left( x, \alpha, \beta, \lambda \right) \longmapsto \left( -x, - \alpha, - \beta, - \lambda \right)$.
In other words, the phase space is fully described if we take into account only non-negative values of $\lambda$. 
Note that the presence of the coupling does not allow for more symmetries.
Recall that, in order to avoid singularities related to the definition of the disformal function in~\eqref{cdv}, we only consider $\mu \geq 0$.

Summarising, we have a three-dimensional phase space, defined according to

\begin{equation}
-1\leq x \leq1\ \wedge \  0\leq y \leq\sqrt{1-x^2} \ \wedge \ 0\leq \sigma \leq \frac{1}{6 x^2},
\label{exisdd}
\end{equation}

\noindent and four free parameters, $\alpha, \beta \in \mathds{R}$ and $\lambda, \mu \in \mathds{R}_{\geq 0}$, to fully characterise the system.

\subsection{Dynamical System Analysis} \label{sec:disfdynanal}

The fixed points with an arbitrary constant equation of state parameter, $\gamma$, are found by setting the left-hand side of the autonomous equations~\eqref{xld},~\eqref{yld} and~\eqref{sild} equal to zero and solving the resulting polynomial equations for $x$, $y$ and $\sigma$. In what follows, we also include the study of the set of fixed points at the compactified plane $\Sigma = \pi/2$.

For a thorough analysis we examine the relevant cosmological parameters, $\Omega_{\phi}$, $Z$, $w_{\phi}$ and $w_{\rm eff}$, as defined in Eq.~\eqref{omegad},~\eqref{zd},~\eqref{wd} and~\eqref{weffd}, respectively. Evaluating the effective EoS parameter we are able to present the range of parameters which render accelerated expansion for each fixed point, \textit{i.e.}, which satisfy $w_\text{eff} <-1/3$. 

We are also interested in performing a stability analysis and, whenever possible, we do it through the study of the eigenvalues $\left( e_1,e_2,e_3 \right)$ of the stability matrix $\mathcal{M}$, constructed by consideration of small perturbations around each fixed point (the eigenvalues are listed in Appendix \ref{app:eigdr}). Otherwise, we present the stability study as the result of a thorough numerical investigation.
In general, the stability character of each fixed point has a clear dependence on the parameters. By allowing the disformal function to depend on the kinetic term we are assuming a more general transformation. As we will see, this reflects on the fact that the parameter $\mu$, associated to this generalisation, can be used to widen/narrow the region of the parameter space that renders the fixed points stable (in comparison with the $\mu = 0$ case, \cite{nelson}).

The restriction to the phase space imposed in Eq.~\eqref{exiss} implies that each fixed point which is a potential attractor of the system must lie inside the region enclosed by this condition. Moreover, the only physically allowed solutions are the ones which respect this constraint throughout the entire evolution. 
In principle it would not be possible to ensure this behaviour \textit{a priori} for all initial conditions. However, for particular choices of $\mu$, we will see further on, that whenever the attractors of the system are physical, initial conditions inside the physical space remain inside it for all times.

In what follows we will consider the concrete cases of pressureless, $\gamma=1$ (labelled as $\rm d$), and relativistic, $\gamma=4/3$ (labelled as $\rm r$), effective fluids, which convey, respectively, an approximation to the late time and early time evolution of the Universe under this model. 

%Furthermore, according to the definition of the variables in~\eqref{variablesd} and the choice for the form of the potential and the coupling functions in~\eqref{cdv}, this system can easily be reduced to previously studied cases, on which we will base our analysis:
%
%\begin{itemize}
%\item If $\mu=0$ the system reduces to disformally coupled quintessence, in the case where the conformal and disformal functions can only depend on the field itself. This was studied in \cite{nelson}.
%
%\item If $\mu=0$ and $\beta \rightarrow - \infty$ this system coincides with the conformally coupled quintessence case. This was addressed in \cite{PhysRevD.62.043511, Gumjudpai:2005ry, Bahamonde:2017ize}.
%
%\item Finally, if $\mu=0$, $\beta \rightarrow - \infty$ and $\alpha=0$ we are left with the standard uncoupled quintessence scenario, presented in \cite{Copeland:1997et, Bahamonde:2017ize}.
%
%\end{itemize}

%In Appendix \ref{app:mu0} we present the fixed points for the system where the disformal function only depends on the field, corresponding to setting $\mu=0$ in equations~\eqref{xld},~\eqref{yld} and~\eqref{sild}, as described above. The fixed points listed correspond to a fluid with an arbitrary constant equation of state parameter $\gamma$. We do so in order to make the comparison between the two systems easier. 

\subsubsection{Fixed points, Stability and Phenomenology for a pressureless fluid} \label{sec:disfdust}

First, we take into account the possibility of having a non-relativistic fluid ($\gamma=1$) disformally coupled to the dark energy fluid. The fixed points are labelled (A)$_{\rm d}$-(F)$_{\rm d}$ and are registered in Table \ref{table:gamma1d}. 
Additionally, at the bottom of Table \ref{table:gamma1d}, we register the singular fixed points found on the $\Sigma \rightarrow \pi/2$ plane, inside the physical phase space, \textit{i.e.}, on the $x=0$ line, labelled as (I$_1$)$_{\rm d}$ and (I$_2$)$_{\rm d}$. 

\begin{table*}[t]
\centering
\begin{tabular}{c c c c c c c c c}
\hline\hline \\[-1.5ex]
Name & $x$ & $y$ & $\sigma$  & $\Omega_{\phi}$ & $Z$ & $w_{\phi}$ & $w_{\rm eff}$ & Acceleration \\ [0.5ex] % inserts table %heading
\hline \\[-1.5ex]
(A$_\pm$)$_{\rm d}$ & $\pm 1$ & $0$ & $0$& $1$ & $1$ & $1$ & $1$ & No\\
(B)$_{\rm d}$ & $- \sqrt{\frac{2}{3}} \alpha $ & $0$ & $0$& $ \frac{2}{3} \alpha^2 $ & $1$ & $1$ & $ \frac{2}{3} \alpha^2$ & No  \\
(C)$_{\rm d}$ & $\frac{\lambda}{\sqrt{6}}$ & $\sqrt{1- \frac{\lambda^2}{6}}$ & $0$ & $1$ & $1$ & $\frac{\lambda^2}{3} -1$ & $\frac{\lambda^2}{3} -1$ & $ \lambda^2 < 2$ \\
(D)$_{\rm d}$ & $\frac{\sqrt{\frac{3}{2}} }{\alpha + \lambda}$ & $\frac{\sqrt{\frac{3}{2}+ \alpha^2 + \alpha \lambda}}{\sqrt{(\alpha + \lambda)^2}}$ & $0$& $\frac{\alpha^2 + \alpha \lambda +3}{(\alpha + \lambda)^2}$ & $1$ & $- \frac{\alpha (\alpha + \lambda)}{\alpha^2 + \alpha \lambda+3}$ & $- \frac{\alpha}{\alpha + \lambda}$ & $\alpha  > \frac{\lambda}{2}$ \\
(E$_\pm$)$_{\rm d}$ & $\frac{\sqrt{2} \beta-\sqrt{2 \beta^2 - 3  \left( 1+\mu \right)^2}}{\sqrt{3} \left( 1+\mu \right)}$ & $0$ & $\sigma_{E_\pm}$  & $x_E^2$ & $\sqrt{1-6 \sigma_{E_\pm} x_E^2}$ &  $1$ & $x_E^2$ & No\\
(F$_\pm$)$_{\rm d}$ & $\frac{\sqrt{2} \beta+\sqrt{2 \beta^2 - 3 \left( 1+\mu \right)^2}}{\sqrt{3} \left( 1+\mu \right)}$ & $0$ &  $\sigma_{F_\pm}$ & $x_F^2$ & $\sqrt{1-6 \sigma_{F_\pm} x_F^2}$ & $1$ & $x_F^2$  & No\\[1.5ex]
\hline\hline \\[-1.5ex]
(I$_1$)$_{\rm d}$ & $0$ & $0$ & $+ \infty$ & $0$ & $-$ & $-$ & $0$ & No\\
(I$_2$)$_{\rm d}$ & $0$ & $1$ & $+ \infty$ & $1$ & $-$ & $ -1$ & $-1$ & Yes \\[1ex]
\hline
\end{tabular}
\caption{Fixed points of the compactified system of equations~\eqref{xld}-\eqref{sild} for the $\gamma=1$ case (labelled as $\rm d$) and corresponding cosmological parameters as defined in Eqs.~\eqref{omegad},~\eqref{zd},~\eqref{wd} and~\eqref{weffd}. The parameter values that lead to an accelerated expansion of the Universe, \textit{i.e.}, $w_{\text{eff}} < -1/3$, for each fixed point, are also listed. For the fixed points (E$_{\pm}$)$_{\rm d}$ and (F$_{\pm}$)$_{\rm d}$, the expressions $\sigma_{E/F_{\pm}}$ represent the solutions of the polynomial for $\sigma$ given Eq.~\eqref{pol}.}
\label{table:gamma1d}
\end{table*}

The pairs $\sigma_{E_\pm}$ and $\sigma_{F_\pm}$ represent the solutions of the following second order polynomial for the corresponding value of $x_E \equiv x\left[ \left({\rm E}_\pm \right)_{\rm d} \right]$ and $x_F \equiv x \left[ \left({\rm F}_\pm \right)_{\rm d} \right]$, respectively, listed in Table \ref{table:gamma1d}:

\begin{gather}
18 x^3 \mu \left[ 3 x^2 \left( 3 \mu +1 \right) +2 \sqrt{6} x\left( \alpha +\beta \right) -3 \left( \mu +1 \right)  \right]  \sigma^2 - 3 x \left[ 3  x^2  \left( 2 \mu ^2 +9 \mu  +3 \right)+ 2 \sqrt{6} x \left(\alpha -\beta+ \alpha  \mu -2 \beta  \mu  \right)  \right. \nonumber \\
\left. +3 \left( \mu +1 \right) \left( 2 \mu +1 \right)  \right] \sigma + \sqrt{6} \alpha +3 x =0. \label{pol}
\end{gather}

% (the polynomial is given in Appendix \ref{app:s0d}). 
 \noindent This choice of representation for the fixed points is ascribed to the intricate form of the analytical solutions obtained for $\sigma$, when solving the second order equation~\eqref{pol}.

In Table \ref{table:gamma1d} the corresponding cosmological parameters are also listed. According to Eq.~\eqref{weffd}, the fixed points that correspond to the accelerated expansion of the Universe can exist only when $y \neq 0$. Without any further information, by inspection of Table \ref{table:gamma1d}, this restriction implies that the only fixed points capable of providing an accelerated expanding description are (C)$_{\rm d}$ and (D)$_{\rm d}$.

Summing up the information on the existence and stability of the whole set of fixed points, we have:

\begin{enumerate}[label=(\roman*)]

\item Points (A$_\pm$)$_{\rm d}$ are independent from the parameters and so they are always present in the phase space. Since $x$ is the only non-zero dynamical variable, we speak of \textit{scalar field kinetic dominated} solutions. Based on Eq.~\eqref{omegad}, they are also referred to as kination fixed points, as they represent dominance of kinetic energy over potential energy. They are characterised by a stiff equation of state for the field, $w_{\phi}=1$, and, as expected, as $Z=1$ these fixed points present no metric singularity and can be treated under this formalism. They also feature a constant effective EoS, $w_{\rm eff}=1$, and so are never capable of describing an accelerating behaviour.
Depending on the value of the parameters $\alpha$, $\beta$, $\lambda$ and $\mu$, they can be attractors, repellers or saddle points. Point (A$_+$)$_{\rm d}$ is an attractor for

\begin{equation}
\ \ \ \alpha < - \sqrt{\frac{3}{2}}\ \wedge\ \lambda >  \sqrt{6}\ \wedge\ \beta<  \sqrt{\frac{3}{2}} \mathopen{} \left( 1+\mu \right) \mathclose{},
\end{equation}

\noindent and a repeller for

\begin{equation}
\ \ \ \alpha > - \sqrt{\frac{3}{2}}\ \wedge \  \lambda < \sqrt{6}\ \wedge \ \beta > \sqrt{\frac{3}{2}} \mathopen{} \left( 1+\mu \right) \mathclose{}.
\end{equation}

\noindent On the other hand, (A$_-$)$_{\rm d}$ can never be an attractor in this range of parameters and is a repeller for

\begin{equation}
\ \alpha < \sqrt{\frac{3}{2}}\ \wedge\  \lambda \geq 0 \ \wedge \ \beta< - \sqrt{\frac{3}{2}} \mathopen{} \left( 1+\mu \right) \mathclose{}.
\end{equation}

\item The fixed point (B)$_{\rm d}$ corresponds to a scaling solution and is referred to as a \textit{conformal kinetic dominated} solution \cite{10.1143/PTP.40.49}, since the disformal coefficient vanishes at the fixed point, according to $\sigma=0$. It is characterised by a stiff EoS, $w_{\phi}=1$. Its existence translates into a simple condition: $\alpha^2 < 3/2$. The effective EoS is also a function of $\alpha$: $w_{\rm eff}= \left( 2/3 \right) \alpha^2$, which means that this fixed point does not feature accelerating behaviour.
Regarding its stability character, it is an attractor for:

\begin{equation}
 -\sqrt{\frac{3}{2}}<\alpha<0\ \wedge\ \lambda > - \frac{3 +2 \alpha^2}{2 \alpha}\  \wedge\ \beta < - \frac{ \left( 3+  2 \alpha^2 \right) \left( 1+\mu \right)}{4 \alpha}, 
\end{equation}
\noindent and a saddle otherwise.

\item Point (C)$_{\rm d}$ is a \textit{conformal scalar field dominated} solution \cite{HALLIWELL1987341, Burd:1988ss}, since $\sigma$ vanishes at the fixed point and $\Omega_\phi=1$. It is defined for $\lambda^2<6$ with $w_{\phi}=\lambda^2 /3 -1$. The effective EoS parameter is also a function of $\lambda$, \textit{via} $w_{\rm eff} = \lambda^2/3-1$, and so, this point represents an accelerating solution whenever $\lambda^2<2$. One can easily conclude that, for parameter values satisfying the following inequality

\begin{equation}
\ \ \ \ \ \ \alpha< \frac{3- \lambda^2}{\lambda} \ \wedge \ 0<\lambda<\sqrt{6} \ \wedge\ \beta< \frac{\lambda}{2} \mathopen{} \left( 1+ \mu \right) \mathclose{}, 
\end{equation}

\noindent it is an attractor and otherwise it is a saddle point. 

\item Point (D)$_{\rm d}$ is a \textit{conformal scaling} fixed point \cite{Amendola:1999qq, PhysRevD.62.043511,Holden:1999hm,PhysRevD.62.043511} as both $w_\phi$ and $\Omega_\phi$ depend on the parameters and the disformal function vanishes at the fixed point ($\sigma=0$). For its existence to be verified, two inequalities have to be satisfied: $\alpha \left( \alpha + \lambda \right) \geq -3/2$ and $ \lambda \left( \lambda+\alpha \right) \geq 3$. Its effective EoS parameter is a function of $\alpha$ and $\lambda$, and so, this fixed point is capable of describing an accelerated expanding scenario for $\alpha>\lambda/2$. It is found to be stable when

\begin{equation}
\ \ \ \ \ \ \ \ \alpha \left( \alpha + \lambda \right) > -3/2 \ \wedge\ \lambda (\lambda+\alpha) > 3 \  \wedge\ \beta < \frac{\lambda}{2} \mathopen{} \left( 1 + \mu \right) \mathclose{},
\label{fpd}
\end{equation}

 \noindent and a saddle otherwise. 

\item Points (E$_\pm$)$_{\rm d}$ and (F$_\pm$)$_{\rm d}$ are referred to as \textit{disformal} fixed points since they are characterised by a non-vanishing disformal strength, $\sigma\neq0$. These are new fixed points that emerge under the formalism considered here and are a consistent generalisation of the ones found in \cite{nelson}.
All four points present a stiff EoS for the field, $w_\phi=1$, and a possible metric singularity is present according to the role of $Z$ in the Jacobian $J$, Eq. \eqref{zd}, which can be avoided through a proper choice of the parameters. These points are characterised by $y=0$ and, hence, are not capable of describing an accelerated expanding Universe. Due to the intricate dependence of the disformal fixed points on the parameters, we perform an appropriate study to determine the conditions under which the disformal fixed points are inside the physical phase space. 
We started by conjecturing that the boundaries for the existence and stability parameter regions are a generalisation of the ones found for the case of $\mu=0$, with a simple dependence on the parameters $\beta$ and $(1+\mu)$, as found for the other fixed points. We verified this hypotheses analytically for $\mu=1,2,3$. Finally, we were able to confirm numerically that the conjectured parameter regions hold true for general $\mu$ and are given by
%We start by formulating conjectures based on generalisations of the $\mu=0$ case which we later confirm through numerical simulations. Our approach warrants that

\begin{equation}
\beta \geq \sqrt{\frac{3}{2}} \left( 1+ \mu \right)\  \wedge\ \mathopen{}  \Bigg\lbrace \left( \alpha \geq - \sqrt{\frac{3}{2}} x_E \right)\ \vee\  \Bigg[ \mu \geq 1\ \wedge\ \alpha \leq  \left( 1+2\mu \right) \left( \sqrt{\frac{3}{2}}x_E- \frac{2 \beta}{1+\mu} \right) \Bigg] \Bigg\rbrace \mathclose{}, 
\end{equation}
 
 \noindent and 
 
\begin{equation}
\beta \geq \sqrt{\frac{3}{2}} \left( 1+ \mu \right)\  \wedge\  \mu \geq 1\  \wedge\ \alpha \geq  (1+2\mu) \mathopen{}\left(  \sqrt{\frac{3}{2}}x_E- \frac{2 \beta}{1+\mu} \right) \mathclose{}, 
\end{equation} 

\noindent for points (E$_-$)$_{\rm d}$ and (E$_+$)$_{\rm d}$, respectively, and 

\begin{equation}
\beta \leq -\sqrt{\frac{3}{2}} \left( 1+ \mu \right)\  \wedge\ \mathopen{}  \Bigg\lbrace \left( \alpha \leq - \sqrt{\frac{3}{2}} x_F \right)\  \vee\  \Bigg[ \mu \geq 1\ \wedge\ \alpha \geq  (1+2\mu) \left(  \sqrt{\frac{3}{2}}x_F- \frac{2 \beta}{1+\mu} \right) \Bigg] \Bigg\rbrace \mathclose{}, 
\end{equation}

 \noindent and 
 
 \begin{equation}
\beta \leq - \sqrt{\frac{3}{2}} \left( 1+ \mu \right)\  \wedge\  \mu \geq 1\   \wedge\  \alpha \leq  (1+2\mu) \mathopen{} \left(  \sqrt{\frac{3}{2}}x_F- \frac{2 \beta}{1+\mu} \right) \mathclose{},
\end{equation}
 
\noindent for points (F$_-$)$_{\rm d}$ and (F$_+$)$_{\rm d}$, respectively.

The same argument used above yields the following conditions for the stability: 

\begin{equation}
\lambda > \frac{2 \beta}{1+\mu}\ \wedge\ \mu \geq 1\ \wedge\  \alpha \leq  \left( 1+2\mu \right) \mathopen{} \left( \sqrt{\frac{3}{2}}x_E- \frac{2 \beta}{1+\mu} \right) \mathclose{},
\end{equation}

\noindent and

\begin{equation}
\lambda > \frac{2 \beta}{1+\mu},
\end{equation}

\noindent for (E$_-$)$_{\rm d}$ and (E$_+$)$_{\rm d}$, respectively. They are saddles otherwise. The points (F$_-$)$_{\rm d}$ and (F$_+$)$_{\rm d}$ are always saddle points.
In some range of the parameters, when $\mu\geq 1$, the stability regions of (E$_\pm$)$_{\rm d}$ are found to overlap with the ones for the other fixed points. When that is the case, the evolution of the dynamics of the system toward the attractor will depend on the initial conditions.

\item Point (I$_1$)$_{\rm d}$ is a \textit{trivial infinite disformal} fixed point which is always present in the phase space independently of the value of the parameters. It is always a repeller.

\item Point (I$_2$)$_{\rm d}$ is a \textit{potential dominated infinite disformal} fixed point which always exists and is either a saddle point or a repeller.

\end{enumerate}

 It is interesting to highlight that the introduction of the parameter $\mu$, associated to the extension of the disformally coupled system studied in \cite{Sakstein:2014aca, nelson}, manifests itself as a generalisation of the existence/stability parameter regions of each fixed point solution. This corroborates the hypotheses that the introduction of a kinetic dependence on the disformal function can be used to widen the stability parameter region for a fixed point with physical interest at late times.

\begin{figure*}[t]
       \subfloat[$\mu=0$]{\includegraphics[width=0.45\linewidth]{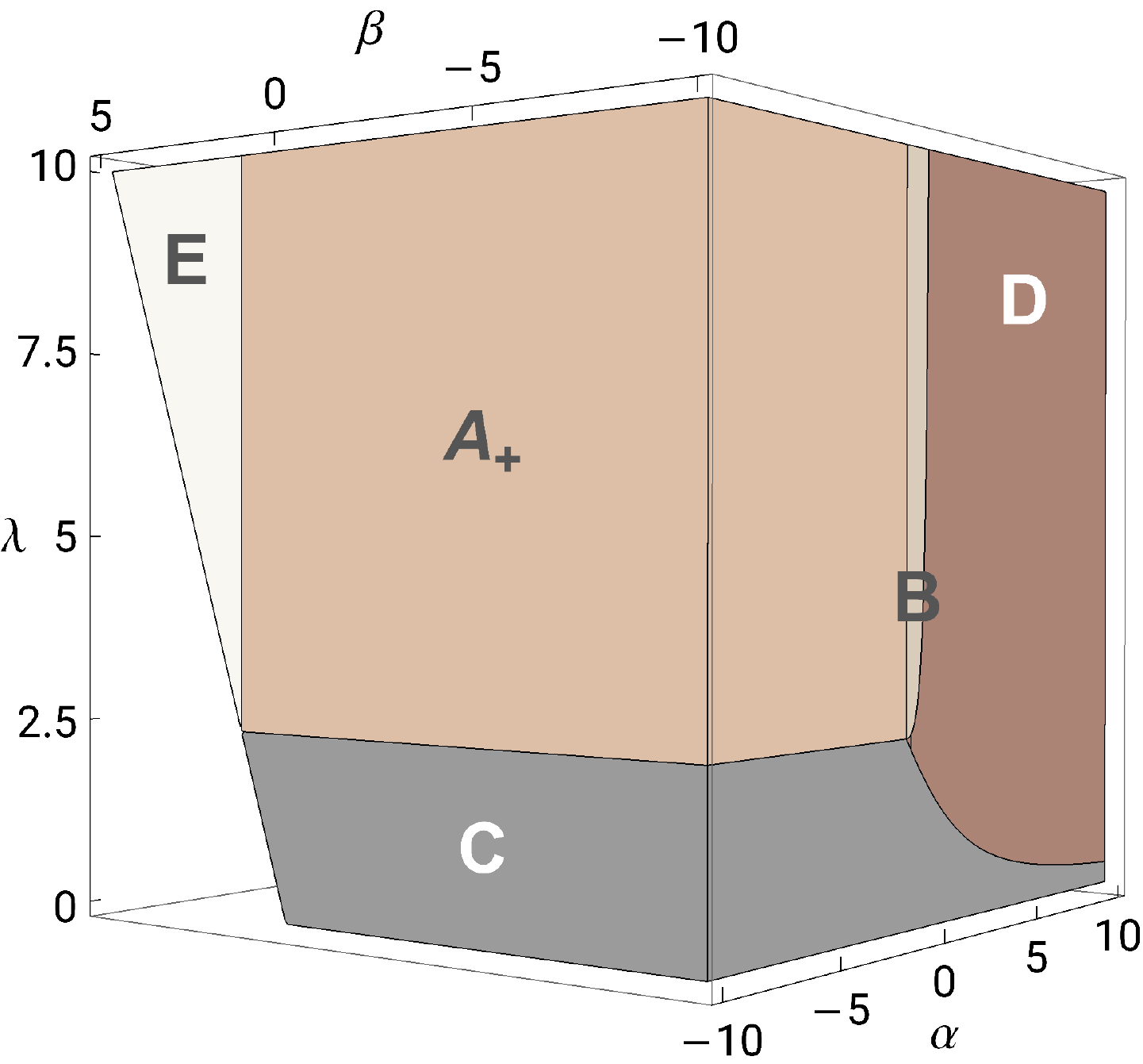}} 
    \hfill
  %       \subfloat[$\mu=0$]{\includegraphics[width=0.3\linewidth]{upupye2}} 
%    \hfill
      \subfloat[$\alpha=1$, $\beta=3$, $\lambda=5$, $\mu=1.3$]{\includegraphics[width=0.38\linewidth]{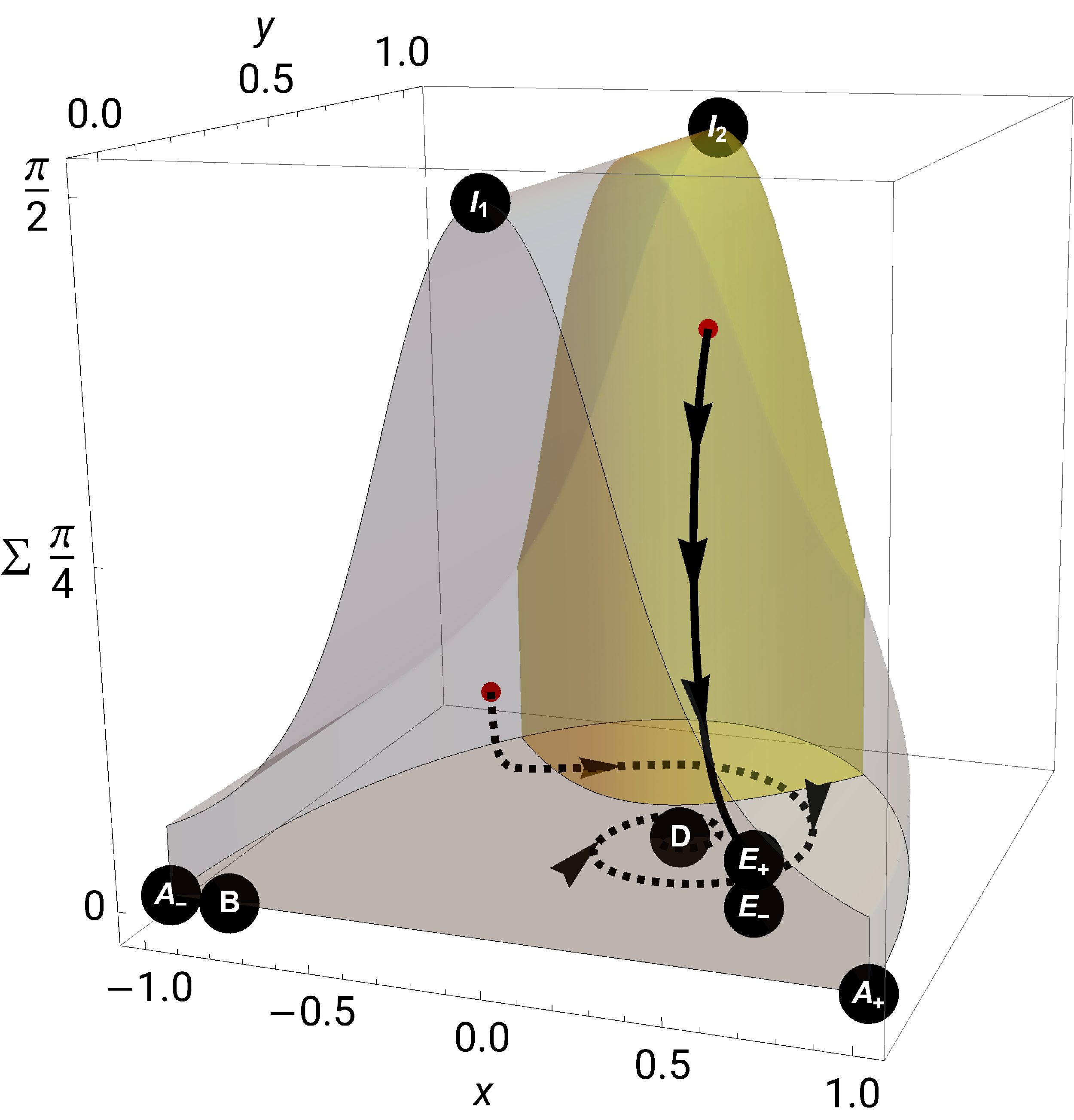}}
  \caption{Panel (a): Illustration of the parameter regions where each fixed point is an attractor of the disformally coupled system with $\mu=0$. The coloured regions are totally disjoint, meaning that, for specific values of the parameters, there is, at most, one attracting fixed point in the phase space. The blank region corresponds to the forbidden set of parameters. Panel (b): Phase portrait of the disformally coupled system for different initial conditions. The coloured regions correspond to the total physical phase space, Eq.~\eqref{exisdd}. The region where the Universe undergoes accelerated expansion, Eq.~\eqref{weffd}, is coloured in yellow. Two different trajectories are drawn (solid and dotted lines), to highlight the fact that the parametric regions for which the fixed points (E$_{\pm}$)$_{\rm d}$ have an attractive character, are allowed to overlap with the stability regions for the other fixed points.}
  \label{fig:ef}
\end{figure*}

When $\mu=0$, the stability parameter regions for the regular fixed points are totally disjoint, meaning that, for a specific choice of the parameters, there is one and only one attracting fixed point in the system, as depicted in Fig. \ref{fig:ef} (a), where colours denote regular attractors and blank spaces stand for forbidden parameter regions. In the latter there are orbits which leave the physical phase space in finite time, leading to non-viable cosmological scenarios.

When $\mu > 1$, we find overlapping stability regions, in which the final cosmological evolution strongly depends on the initial conditions. When that is the case, initial conditions in the basin of attraction of each fixed point may lead to different, but still viable, cosmological evolutions. One example of such a scenario is illustrated in Fig. \ref{fig:ef} (b), where the competing attractors are the fixed points (D)$_{\rm d}$ and (E$_+$)$_{\rm d}$.

In previous works \cite{Sakstein:2014aca, nelson}, the interest of taking disformal couplings has been questioned on the basis that the viable models of disformal dark energy have late-time properties that are equivalent to those found in models with no disformal coupling. This is related to the fact that the disformal fixed points do not have accelerating properties. Additionally, they are characterised by a metric singularity (when $\mu=0$) and this property reduces the conditions under which these fixed points are cosmologically interesting.
However, in these works, the study is restricted to the $D \equiv D ( \phi )$ case. We have seen that, by taking the generalisation $D \equiv D ( \phi, X )$ new disformal fixed points emerge with novel features. Albeit still not capable of being the main driver for the acceleration, this ``kinetic" disformal fixed point solutions exhibit well-defined Jordan frame metrics, under certain conditions. This means that the disformal fixed points can be present with transient properties, thus introducing novel and enriched phenomenology.

\begin{figure*}[t]
%       \subfloat[$\mu=0.1$]{\includegraphics[width=0.31\linewidth]{reg0p1}} 
%       \hfill
      \subfloat{\includegraphics[width=0.33\linewidth]{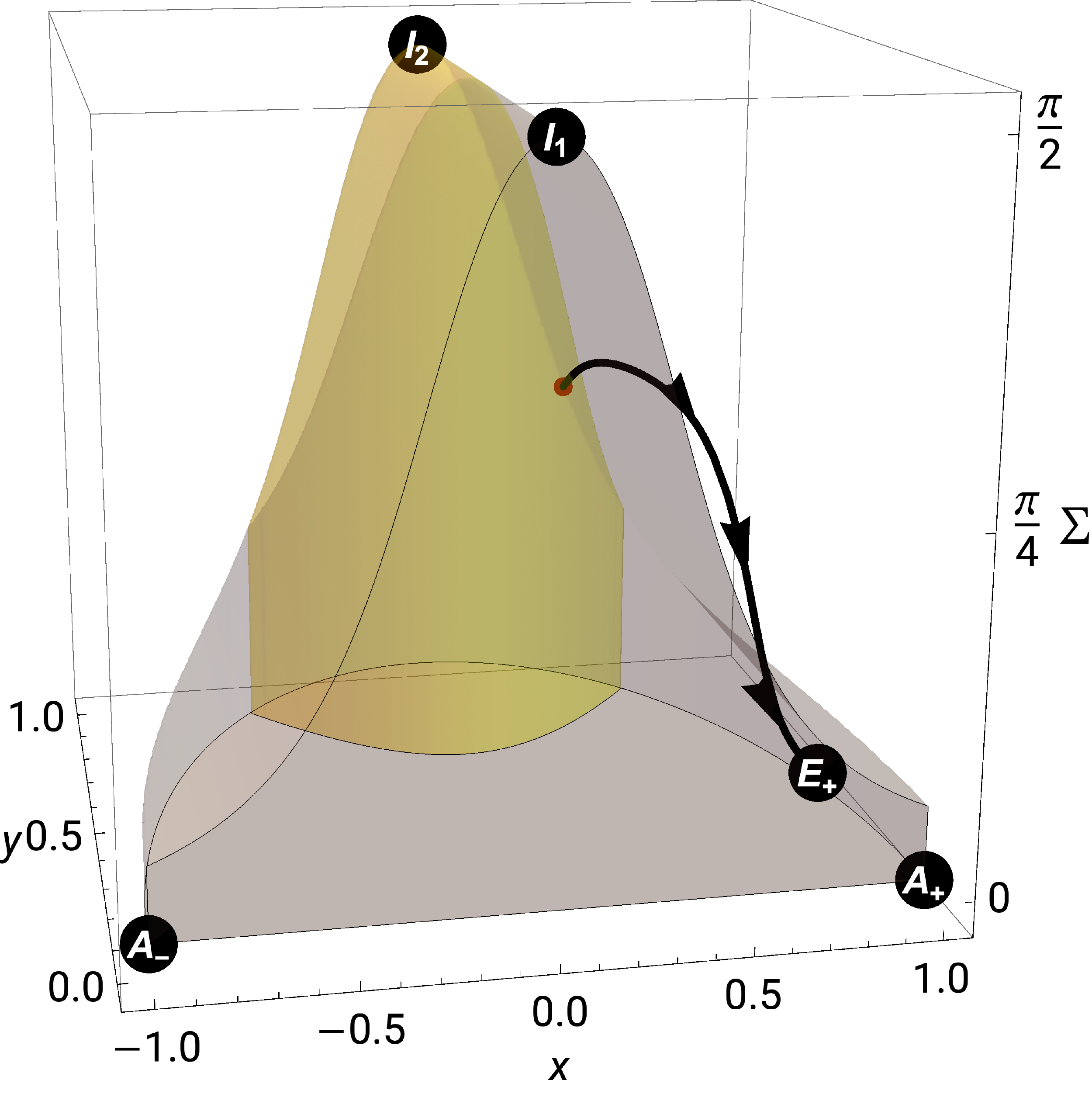}}
                \hfill
      \subfloat{\includegraphics[width=0.5\linewidth]{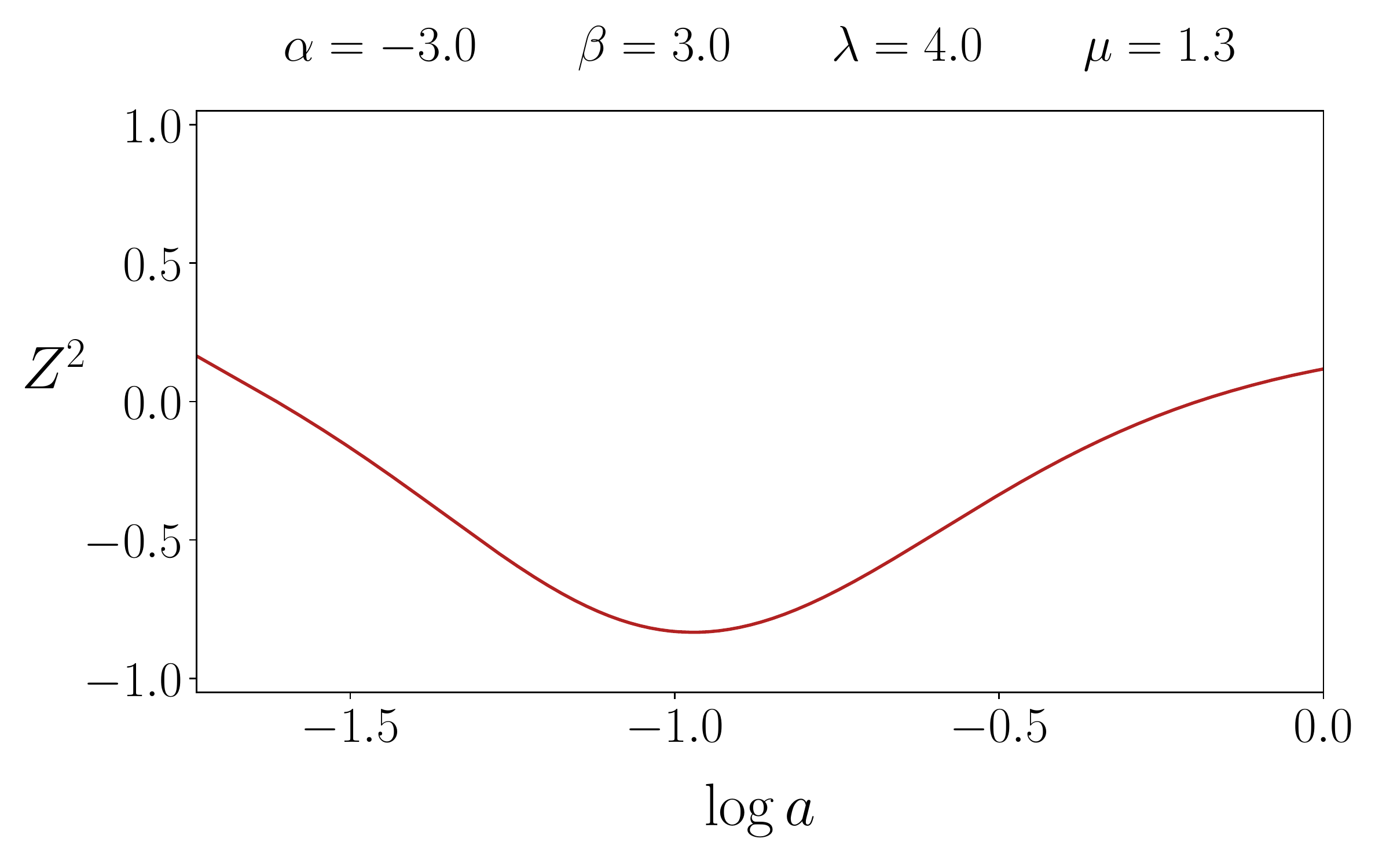}}
  \caption{Example of an orbit which briefly leaves the physical phase space, as defined in~\eqref{exisdd}, in finite time, giving rise to an ill-defined metric transformation.}
  \label{fig:physsp}
\end{figure*}

We identify the presence of a``natural resistance to pathology", as mentioned in \cite{Sakstein:2014aca} (and references therein) for the $\mu=0$ case. In other words, given any set of initial conditions inside the physical phase space, and parameter values such that the attractors are also inside the physical phase space, the orbits will always remain there. We were able to explain this phenomenon under this formalism by evaluating the derivative of the parameter $Z^2$, as defined in~\eqref{zd}, at $Z^2 = 0$, \textit{i.e.}, when we approach the pathological behaviour:

\begin{equation}
\left( Z^2 \right) ' =-6 x \left(x \sigma'+2\sigma x' \right) \bigg|_{\sigma=1/ \left( 6x^2 \right)}= \frac{4 x \mu \left(\mu -1 \right) \left[2 \sqrt{6} \beta  x^2+\sqrt{6} \left( \mu +1 \right) \left(\lambda  y^2-\sqrt{6} x\right)\right]}{\left(y^2-1\right) \left( \mu +1 \right) \left( 3 \mu +1 \right) +\left(1+8\mu-\mu ^2\right) x^2},
\label{zlinha}
\end{equation}

\noindent which vanishes for $\mu=0$. This implies that, as the orbits approach the $Z=0$ surface, they inevitably become trapped and freeze at the singular surface. From Eq.~\eqref{zlinha} we can conclude that this is exceptionally true for the $\mu=1$ case as well. On the other hand, when $\mu \neq \lbrace 0,1\rbrace$, depending on the parameter values and on the initial conditions, there is no longer a mechanism responsible for holding the orbits inside the physical phase space, and so they may momentarily escape. One example where an orbit briefly leaves the physical phase space, and therefore the metric transformation becomes ill-defined, is illustrated in Fig. \ref{fig:physsp}.

Motivated by the discussion above, from now on, we will focus only on the case where the disformal coupling function has a linear dependence on the kinetic term, \textit{i.e.}, we take $\mu=1$. This assumption warrants that the metric transformation is always well-defined (or at the most, it may become singular). Moreover, according to Eq.~\eqref{wtransdr}, the EoS parameter in the Jordan frame identically coincides with the one in the Einstein frame. The latter will be an important simplification for the case of a radiation-like fluid, on which we focus in the following section.

\subsubsection{Fixed points, Stability and Phenomenology for a relativistic fluid} \label{sec:disfrad}

Now, we wish to consider a relativistic fluid ($\gamma = 4/3$) disformally coupled to the dark energy fluid. As a simplification, and driven by the study for the non-relativistic fluid, we focus only on the case of a disformal coupling function which depends linearly on the kinetic term, \textit{i.e.}, we take $\mu=1$, in which case the EoS parameter is still invariant, according to Eq.~\eqref{wtransdr}.

The fixed points for the system with $\gamma=4/3$ and $\mu=1$ are registered in Table \ref{table:gamma43m1}, including the fixed points in the compactified plane $\Sigma= \pi/2$. In a similar fashion to the previous case, we list the relevant cosmological parameters, together with the range of values of the parameters that correspond to accelerated expansion.

\begin{table*}[ht]
\centering
\begin{tabular}{c c c c c c c c c}
\hline\hline \\[-1.5ex]
Name & $x$ & $y$ & $\sigma$  & $\Omega_{\phi}$ & $Z$ &$w_{\phi}$ & $w_{\rm eff}$ & Acceleration \\ [0.5ex] % inserts table %heading
\hline \\[-1.5ex]
(O)$_{\rm r}$ &$0$ & $0$ & $0$ &$0$ & $1$ & $-$ & $1/3$ & No\\
(A$_\pm$)$_{\rm r}$ & $\pm 1$ & $0$ & $0$& $1$ & $1$ & $1$ & $1$ & No\\
(B)$_{\rm r}$ & $\frac{\lambda}{\sqrt{6}}$ & $\sqrt{1- \frac{\lambda^2}{6}}$ & $0$ & $1$ & $1$ & $\frac{\lambda^2}{3} -1$ & $\frac{\lambda^2}{3} -1$ & $0\leq\lambda < 2$\\
(C)$_{\rm r}$ &  $ \sqrt{\frac{2}{3}} \frac{2}{\lambda}$ & $\frac{2}{\sqrt{3\lambda^2} }$ & $0$& $\frac{4}{\lambda^2}$ & $1$ & $ \frac{1}{3}$ & $1/3$ & No\\
(D$_\pm$)$_{\rm r}$ & $\frac{1}{4} \left(\sqrt{6} \beta - \sqrt{6 \beta ^2-32}\right)$ & $0$ & $\sigma_{D_{\pm}}$& $x_{D}^2$ & $\sqrt{1-\sigma_{D_{\pm}}x_{D}^2}$ & $1$ & $x_D^2$ & No \\
(E$_\pm$)$_{\rm r}$ & $\frac{1}{4} \left(\sqrt{6} \beta + \sqrt{6 \beta ^2-32}\right)$ & $0$ & $\sigma_{E_{\pm}}$& $x_{E}^2$ & $\sqrt{1-\sigma_{E_{\pm}}x_{E}^2}$ & $1$ & $x_E^2$ & No \\[1.5ex]
\hline\hline \\[-1.5ex]
(I$_1$)$_{\rm r}$ & $0$ & $0$ & $+ \infty$ & $0$ & $-$ & $-$ & $1/3$ & No\\
(I$_2$)$_{\rm r}$ & $0$ & $1$ & $+ \infty$ & $1$ & $-$ & $ -1$ & $-1$ & Yes \\[1ex]
\hline
\end{tabular}
\caption{Fixed points of the compactified system of equations~\eqref{xld}-\eqref{sild} for the $\gamma=4/3$ and $\mu=1$ case, \textit{i.e.}, a linear dependence on $X$ in the disformal coupling function defined in~\eqref{cdv}. The corresponding cosmological parameters $\Omega_{\phi}$, $Z$, $w_{\phi}$ and $w_{\rm eff}$, as defined in Eqs.~\eqref{omegad},~\eqref{zd},~\eqref{wd} and~\eqref{weffd}, and the parameter values that lead to accelerated expansion of the Universe ($w_{\rm eff} < -1/3$) for each fixed point, are also displayed. For the fixed points (D$_{\pm}$)$_{\rm r}$ and (E$_{\pm}$)$_{\rm r}$, the expressions $\sigma_{D/E_{\pm}}$ stand for the solutions of the second order polynomial for $\sigma$ given Eq.~\eqref{pol2}.}
\label{table:gamma43m1}
\end{table*}

Similarly, the pairs $\sigma_{D_{\pm}}$ and $\sigma_{E_{\pm}}$, presented in Table \ref{table:gamma43m1}, are the solutions of the following second-order polynomial, for the corresponding value of $x$:

\begin{equation}
18 x^2 \left( 4 x^2+\sqrt{6} x \beta -2 \right) \sigma^2 -3 \left[ 14 \left(x^2+1 \right)-3 \sqrt{6} x \beta \right] \sigma +1 =0.
\label{pol2}
\end{equation}

Some comments can be stated regarding the existence and stability of each fixed point in Table \ref{table:gamma43m1}:

\begin{enumerate}[label=(\roman*)]

\item Point (O)$_{\rm r}$ is a complete \textit{radiation dominated} solution. It always exists as it is totally independent of the value of the parameters. It is characterised by an indetermination in the scalar field's EoS parameter, $w_{\phi}$ (this is not physically relevant since the effective energy of the field -- sum of kinetic and potential energy -- is zero) and $\Omega_{\phi}=0$. It stands for a pure radiation dominated fixed point and it is always a saddle point.

\item Points (A$_\pm$)$_{\rm r}$ represent \textit{scalar field kinetic dominated} solutions. They are independent of the introduced parameters and so they always exist. They present a stiff EoS parameter for the field, $w_\phi = 1$ and, as expected, present no metric singularity. The effective EoS parameter is also constant, $ w_{\rm eff} = 1$. They can never be stable and they are repelling nodes in the parameter range

\begin{equation}
\beta >\sqrt{6}\ \wedge \ 0 \leq \lambda \leq \sqrt{6},
\end{equation}

\noindent for (A$_+$)$_{\rm r}$ and 

\begin{equation}
\beta < -\sqrt{6}\ \wedge \  \lambda \geq 0,
\end{equation}

\noindent for (A$_-$)$_{\rm r}$, making them possible past attractors of the system.

\item Regarding the fixed point (B)$_{\rm r}$, it is a \textit{conformal scalar field dominated} solution, which was also found in the case of a pressureless fluid. Similarly, it is only defined for $0 \leq \lambda \leq \sqrt{6}$ and $w_\phi = \lambda^2/3 - 1$. A brief inspection of the eigenvalues shows that it can only be an attractor or a saddle point, since one of the eigenvalues is always negative in the fixed point existence range. It is an attractor for:

\begin{equation}
 0<\lambda <2 \ \wedge \ \beta <\lambda,
\end{equation}

\noindent and a saddle otherwise.

\item Point (C)$_{\rm r}$ is a \textit{scaling} fixed point and can only exist when $\lambda \geq 2$.
It is characterised by a constant EoS parameter for the field, $w_\phi=1/3$ and presents no metric singularity. The effective equation of state parameter is, accordingly, $w_{\rm eff}=1/3$. It is an attractor for 

\begin{equation}
 \lambda >2 \ \wedge \ \beta <\lambda,
\end{equation}

\noindent and a saddle otherwise.

\item Points (D$_\pm$)$_{\rm r}$ and (E$_\pm$)$_{\rm r}$ are \textit{disformal} fixed points. Analogously to the dust-like disformal fixed points, these are novel solutions and, moreover, generalise the ones found in \cite{nelson}. The points (D$_\pm$)$_{\rm r}$ and (E$_\pm$)$_{\rm r}$ can only exist for $\beta \geq \sqrt{6}$ and $\beta \leq - \sqrt{6}$, respectively. They are scaling fixed points in the sense that the density parameter of the field is a function of the parameter $\beta$. The effective EoS is also a function of the parameters. Moreover, the EoS of the field is constant, $w_{\phi}=1$.

In the same fashion as for the dust-like scenario, general linear stability analysis techniques can not be applied in order to obtain simple conditions on the parameters for the stability of the disformal fixed points since these have a complex form and high-order dependence on the parameter $\beta$. Therefore, the stability study was also performed on the basis of conjectures which were subsequently verified numerically. We find that (D$_-$)$_{\rm r}$ and (E$_\pm$)$_{\rm r}$ are always saddle points and moreover, in the range

\begin{equation}
\beta > \sqrt{6} \wedge \lambda > \beta,
\end{equation}

\noindent (D$_+$)$_{\rm r}$ is an attracting solution, otherwise it is a saddle.
The parametric stability region for the fixed point (D$_+$)$_{\rm r}$ is found to always overlap with the one for (C)$_{\rm r}$, emphasizing that the dependence of the final cosmological evolution on the initial conditions is also present for radiation-like fluids.

\item Point (I$_1$)$_{\rm r}$ is a \textit{trivial infinite disformal} fixed point and is always present in the phase space independently of the value of the parameters. It is always a saddle.

\item Point (I$_2$)$_{\rm r}$ is a \textit{potential dominated infinite disformal} fixed point which always exists and is always a repeller.

\end{enumerate}

From this dynamical analysis we gather that the only possible (regular) past attractors of the model with relativistic fluids coupled to the dark energy source are (A$_\pm$)$_{\rm r}$. This is consistent with what was found in previous studies and again, the novel feature introduced by the disformal coupling is the shaping of the stability regions and the emergence of early scaling solution candidates.

For the cosmological analysis of this model, on which we will focus in the following section, we will be concerned with late time cosmologies and, therefore, will not focus on couplings to relativistic fluids. 

\section{Cosmological Analysis}

In this work we seek for viable cosmological trajectories, capable of describing the expansion history of the Universe, in which the kinetic disformal coupling introduces novel features.

Recall that, as a first approximation, we have only considered the case of a single fluid coupled to the dark energy component: a pressureless effective fluid or a relativistic fluid (the complete two-fluid dynamical analysis would depend on 8 parameters). This allows us to have a qualitatively grasp of the dynamics for the two-fluid case, as in the past history of the Universe, the contribution of non-relativistic matter is negligible whereas in the present, it is the contribution of the relativistic fluids which is practically insignificant. Thus, in the asymptotic past, we should look for the fixed points corresponding to a Universe totally dominated by radiation. At late times, however, we wish to look for a combination of the effect of the dark energy and pressureless fluids. Furthermore, the scalar field should only play a relevant role at late times. This means that, at early times, and in order to avoid signatures of early dark energy \cite{Wetterich:2004pv,Aghanim:2018eyx}, the initial values of the variables associated to the scalar field ($x$ and $y$) must be close to zero. 

Regarding the previous analysis, the possible past attractors are (A$_{\pm}$)$_{\rm r}$ and the only possible late-time fixed points, that allow for accelerated solutions, are the scalar field dominated fixed point (C)$_{\rm d}$ and the conformal scaling fixed point (D)$_{\rm d}$. The main difference resides in the nature of the asymptotic behaviour for the solutions (C)$_{\rm d}$ and (D)$_{\rm d}$. The point (C)$_{\rm d}$ stands for a universe completely dominated by the dark energy fluid near the fixed point. When (C)$_{\rm d}$ is the attractor of the system, the coincidence problem, associated to the current observed values of the energy densities, persists. On the other hand, when the attractor is the scaling fixed point (D)$_{\rm d}$, the energy densities of the interacting species scale with each other and remain constant (for a fixed value of $\alpha$ and $\lambda$), independently of the choice for the initial conditions. This could, in principle, alleviate the coincidence problem. 

Recall there are no new candidates for the late time attractor, in comparison with the conformally coupled quintessence scenario \cite{PhysRevD.62.043511,Gumjudpai:2005ry,Bahamonde:2017ize}. This result is a feature inherited from the purely $\phi$-dependent disformal coupling \cite{nelson, Sakstein:2014aca}. 
We will therefore investigate the impact of the disformal coupling in the two possible late time attractor scenarios leading to an accelerated expansion. Throughout this section, we express the expansion history of the Universe in terms of the cosmological redshift $z$, related to the scale factor as $z=1/a-1$.

\begin{figure*}[h]
\includegraphics[width=0.6\linewidth]{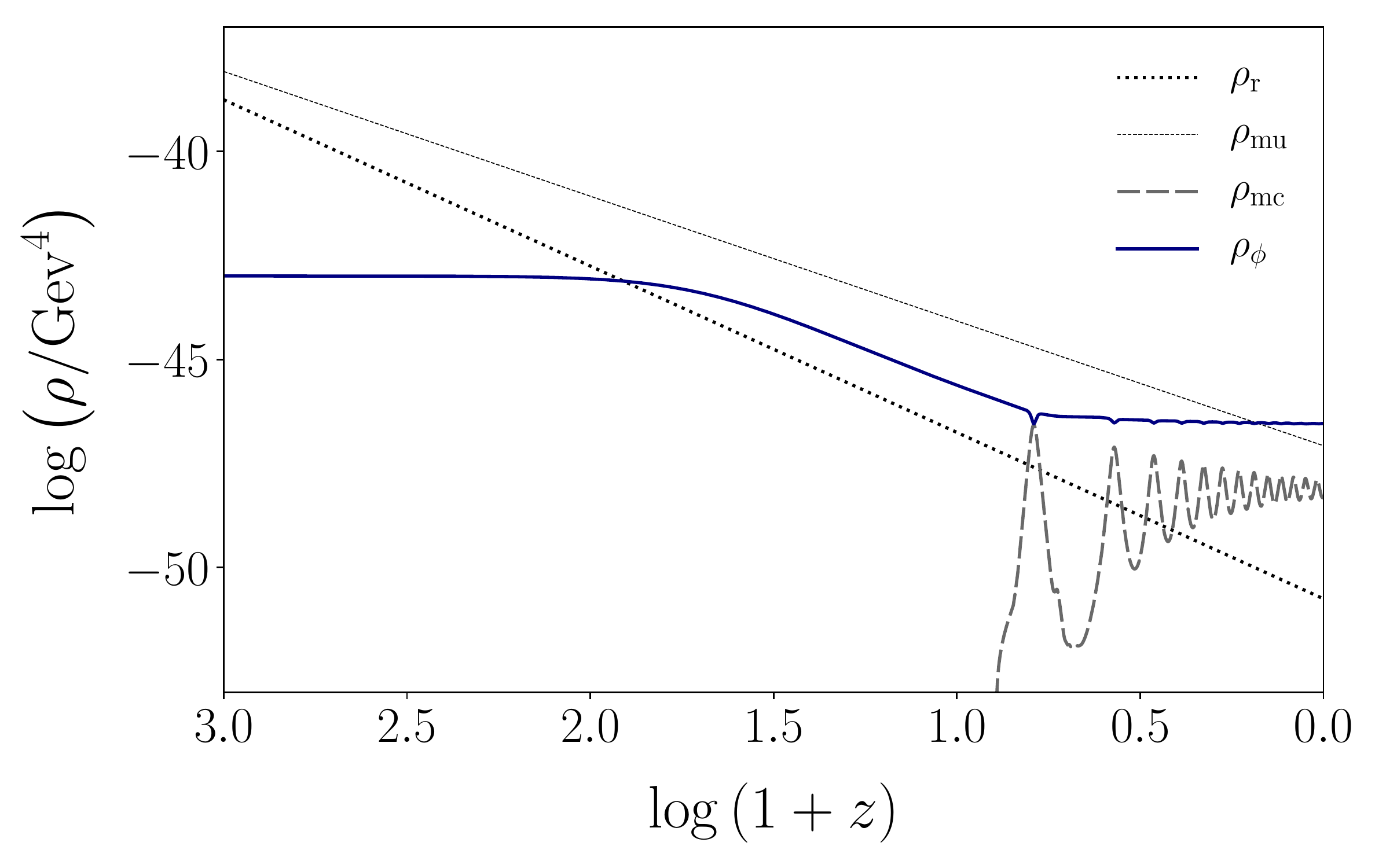}
\caption{\label{fig:densities1}Typical evolution of the energy densities of radiation ($\rho_{\rm r}$), uncoupled matter ($\rho_{\rm mu}$), coupled dark matter ($\rho_{\rm mc}$) and the scalar field ($\rho_{\rm \phi}$), for the cases presented in Table \ref{tabela1}. We have used $\Omega_{\rm mu} = 0.28$ at present and $\lambda = 10$ and $\alpha = 350$.}
  \label{fig:cosm}
\end{figure*}

Considering the scaling solution (D)$_{\rm d}$, the typical evolution of the energy densities is illustrated in Fig.~\ref{fig:densities1}. We studied the three cases specified in Table \ref{tabela1} and confirmed that the energy density evolution remains practically unchanged from case to case. 
\begin{table*}[h!]
\centering
\begin{tabular}{|c|c|c|c|}
\hline %[-1.5ex]
~ & $\beta$ & $\mu$ & $M$ (eV) \\ 
\hline %[0.5ex] % inserts table %heading
(a) &1 & 0 & $10^{-1}$ \\
 \hline 
(b) & 1 & 0 & $3\times 10^{-3}$ \\ 
\hline 
(c) &1 & 1 & $5\times 10^{-9}$ \\ 
\hline  %[-1.5ex]  \\[1ex]
\end{tabular}
\caption{\label{tabela1} The three cases studied for the late time conformal scaling solution with $\lambda = 10$ and $\alpha = 350$.}
\end{table*}

\begin{figure*}[t]
\includegraphics[width=0.495\linewidth]{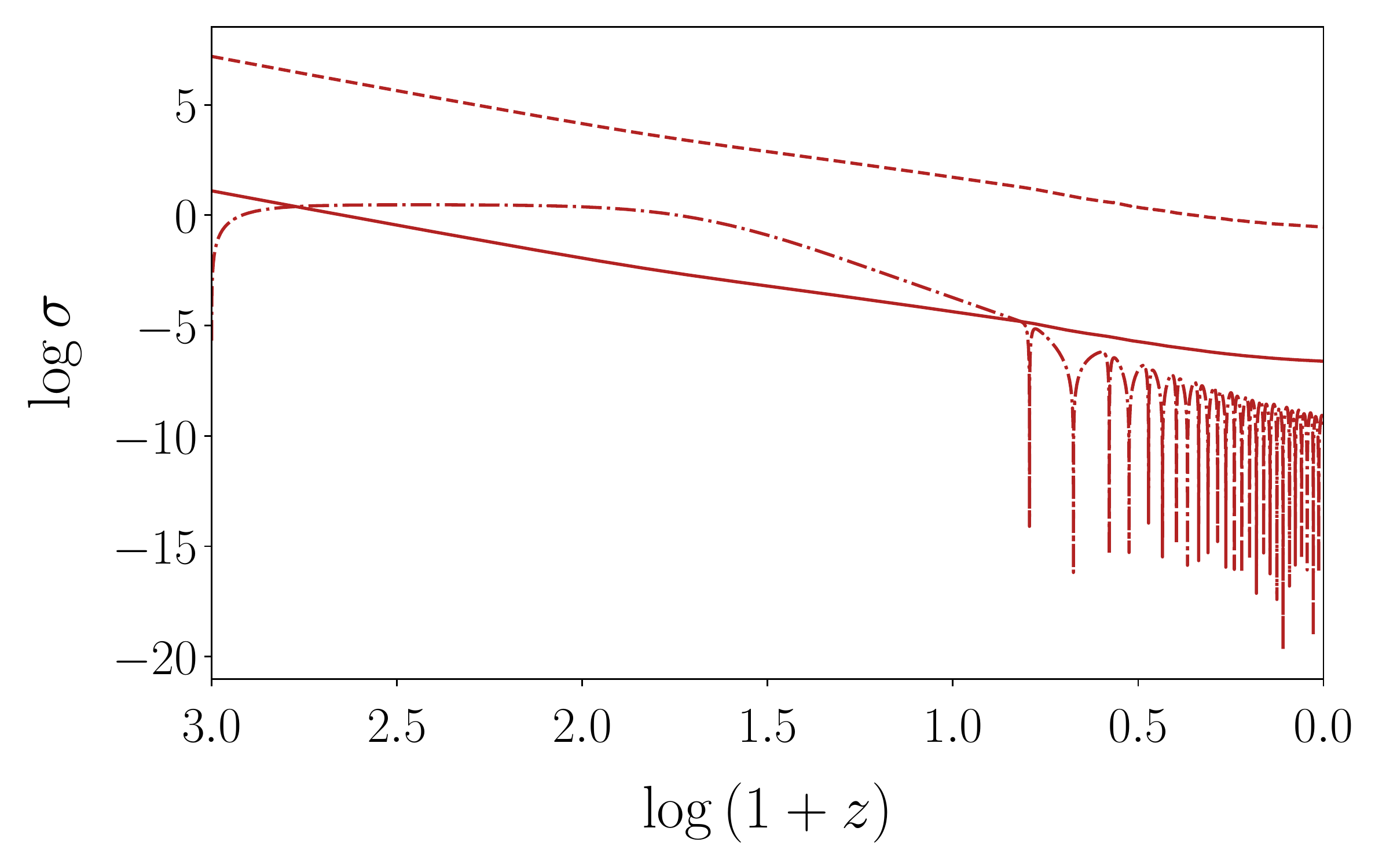}
\includegraphics[width=0.495\linewidth]{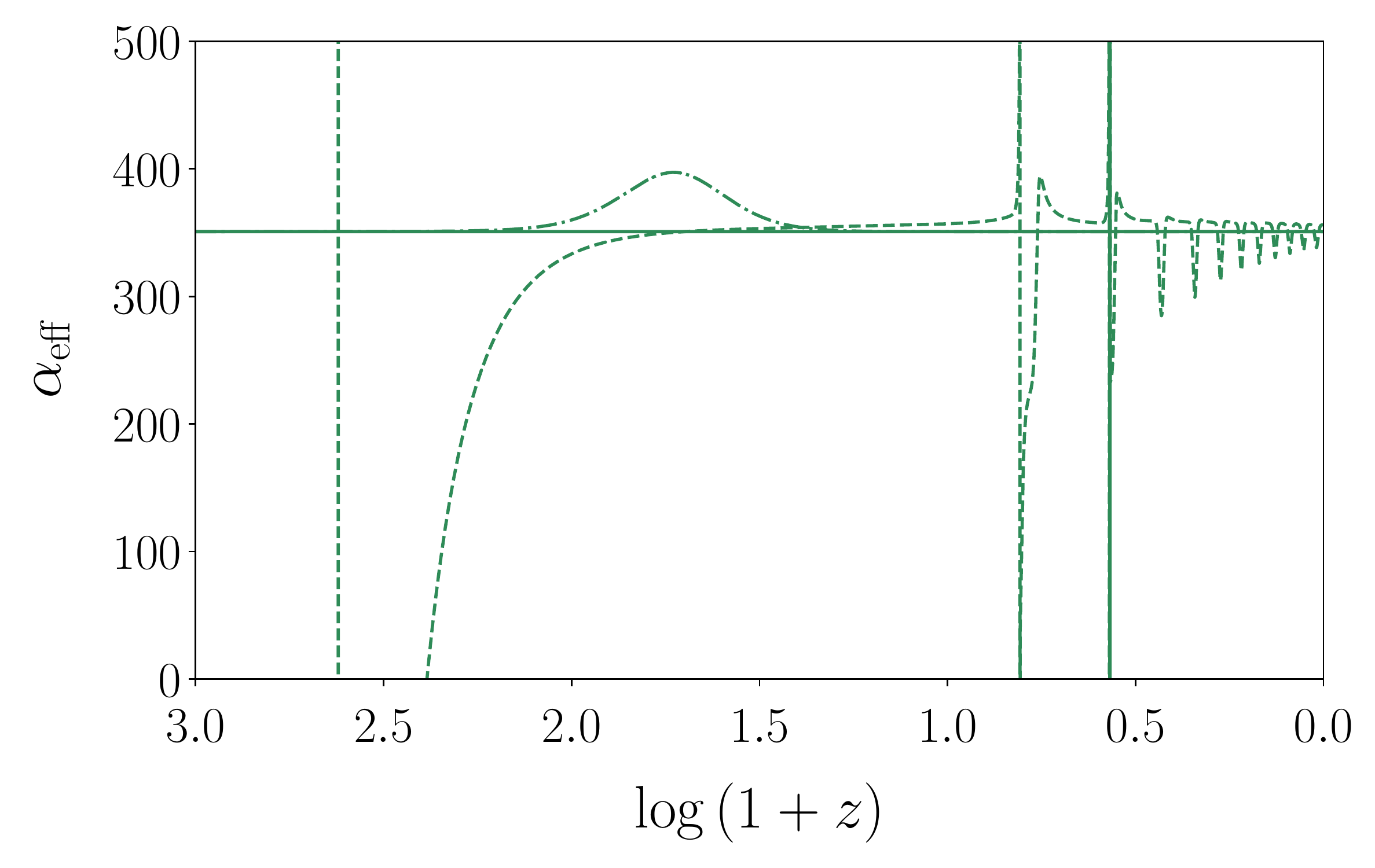}
\caption{\label{fig:sigma1} Left panel, the evolution of $\sigma$ for the models in Table \ref{tabela1}. Model (a), solid line; model (b), dashed line; model (c) dot-dashed line. Right panel, the evolution of the effective conformal coupling $\alpha_{\rm eff}$ for the same models.}
\end{figure*}

This is simple to understand given that at early times it is the scalar potential that rules the dynamics of the scalar field and only when the scaling regime between the field and coupled dark matter is attained the disformal coupling may have any effect. However, as can be seen in Fig.~\ref{fig:sigma1}, $\sigma$ quickly decays and renders the disformal contribution negligible. Even the case with $\mu = 1$ shows an evolution of $\sigma$ that first increases but quickly turns around and decays. 

The importance of the disformal coupling compared to the conformal can also be measured in terms of an effective conformal coupling, defined as
\begin{eqnarray} 
\alpha_{\rm eff} = -Q/\kappa \rho_{\rm mc},
\end{eqnarray}
where $\rho_{mc}$ stands for coupled matter, that is to say, coupled dark matter. 
It can be observed in Fig.~\ref{fig:sigma1} that $\alpha_{\rm eff}$ can have strong departures from the conformal parameter $\alpha$ at early times, but as the disformal coupling decays, all cases tend to approach $\alpha$ at late times.

The extremely large coupling required to obtain a viable scaling solution is known to lead to overgrowth of dark matter perturbations and therefore excluded observationally \cite{amendola2015dark}. We will now turn to the alternative possibility, the scalar field dominated solution, fixed point (C)$_{\rm d}$. Both the coupling and $\lambda$ are expected to be small, and we chose the particular combinations shown in Table \ref{tabela2} in the ballpark of the best fit parameters found in \cite{vandeBruck:2017idm}.
\begin{table*}[h!]
\centering
\begin{tabular}{|c|c|c|c|}
\hline %[-1.5ex]
~& $\beta$ & $\mu$ & $M$ (eV) \\ 
\hline %[0.5ex] % inserts table %heading
(a) & $-\alpha$ & 0 & $1\times 10^9$ \\
 \hline 
(b) & 1 & 0 & $5\times10^{-1}$ \\ 
\hline 
(c) & 1 & 1 & $6\times10^{-8}$ \\ 
\hline
(d) & 1 & 1 & $6\times10^{-17}$ \\ 
\hline  %[-1.5ex]  \\[1ex]
\end{tabular}
\caption{\label{tabela2} The four cases studied for the late time scalar field dominated solution with $\lambda = 1$ and $\alpha = 0.02$.}
\end{table*}

\begin{figure*}[h]
\includegraphics[width=0.6\linewidth]{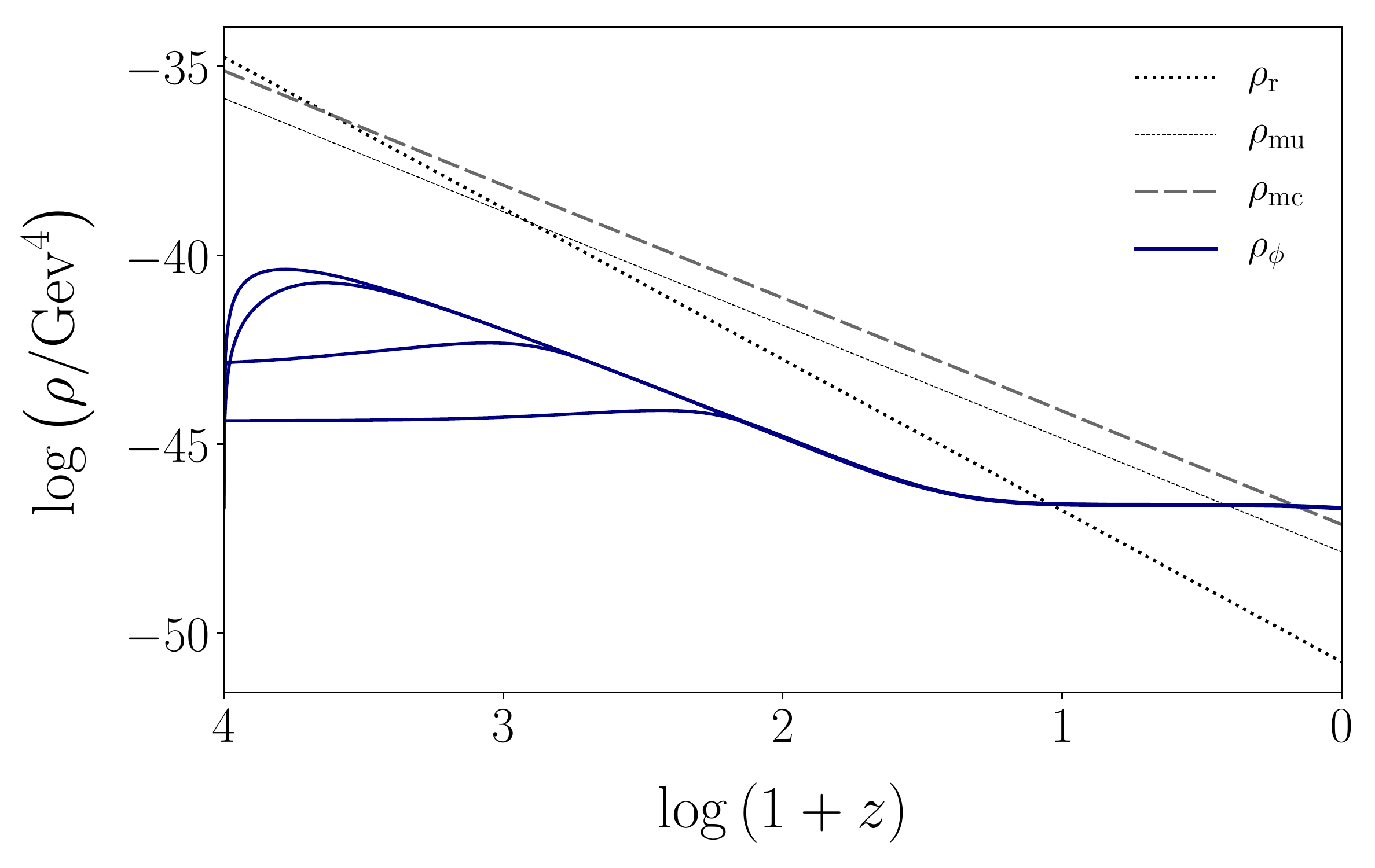}
\caption{\label{fig:densities2} The evolution of the energy densities of radiation ($\rho_{\rm r}$), uncoupled matter ($\rho_{\rm mu}$), coupled dark matter ($\rho_{\rm mc}$) and the scalar field ($\rho_{\rm \phi}$). We have used $\Omega_{\rm mu} = \Omega_b$ and $\lambda=1$, $\alpha = 0.02$. The solid lines represent the evolution of the scalar field energy density within the models (a) to (d) from top to bottom.}
\end{figure*}

\begin{figure*}[h]
\includegraphics[width=0.495\linewidth]{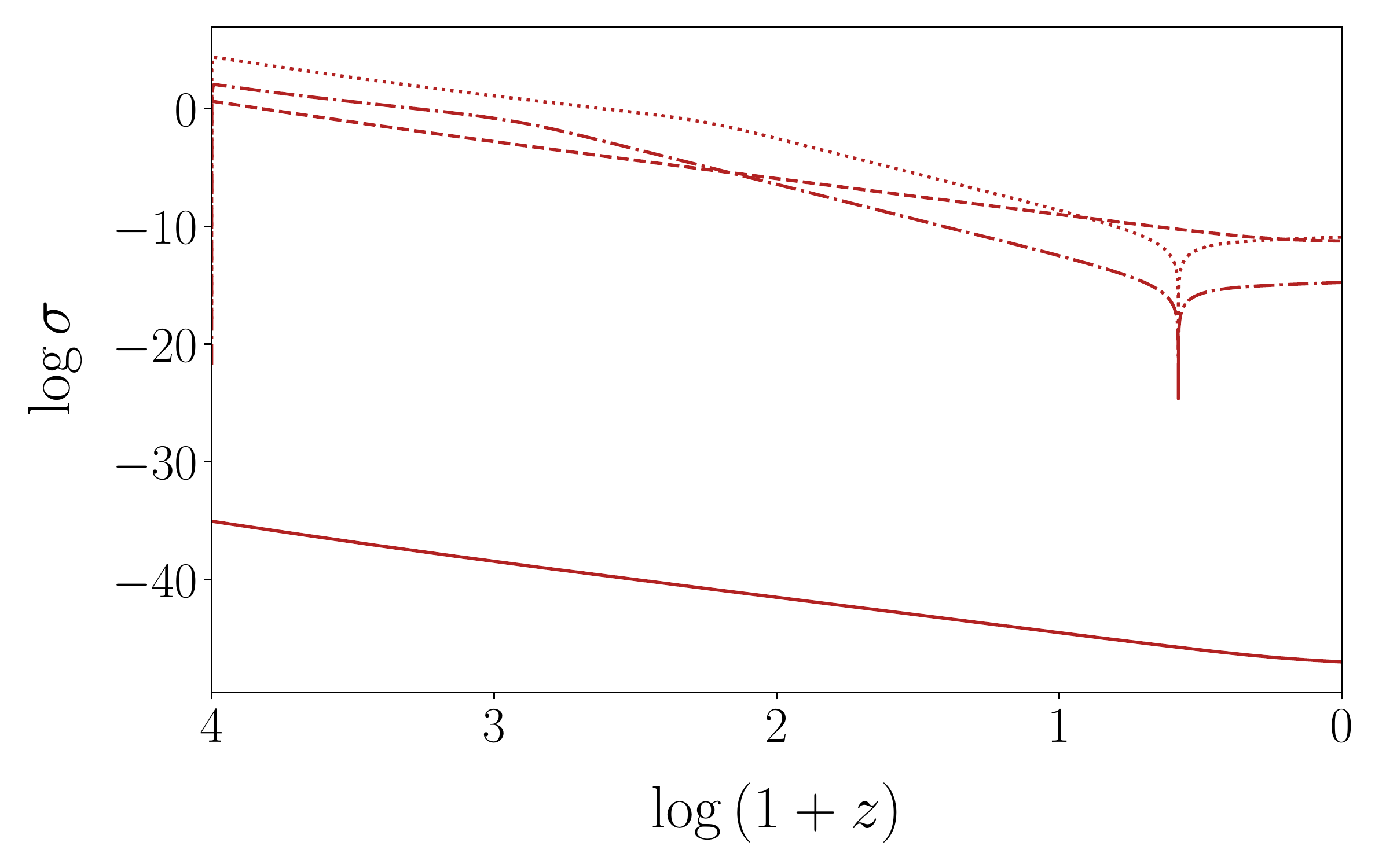}
\includegraphics[width=0.495\linewidth]{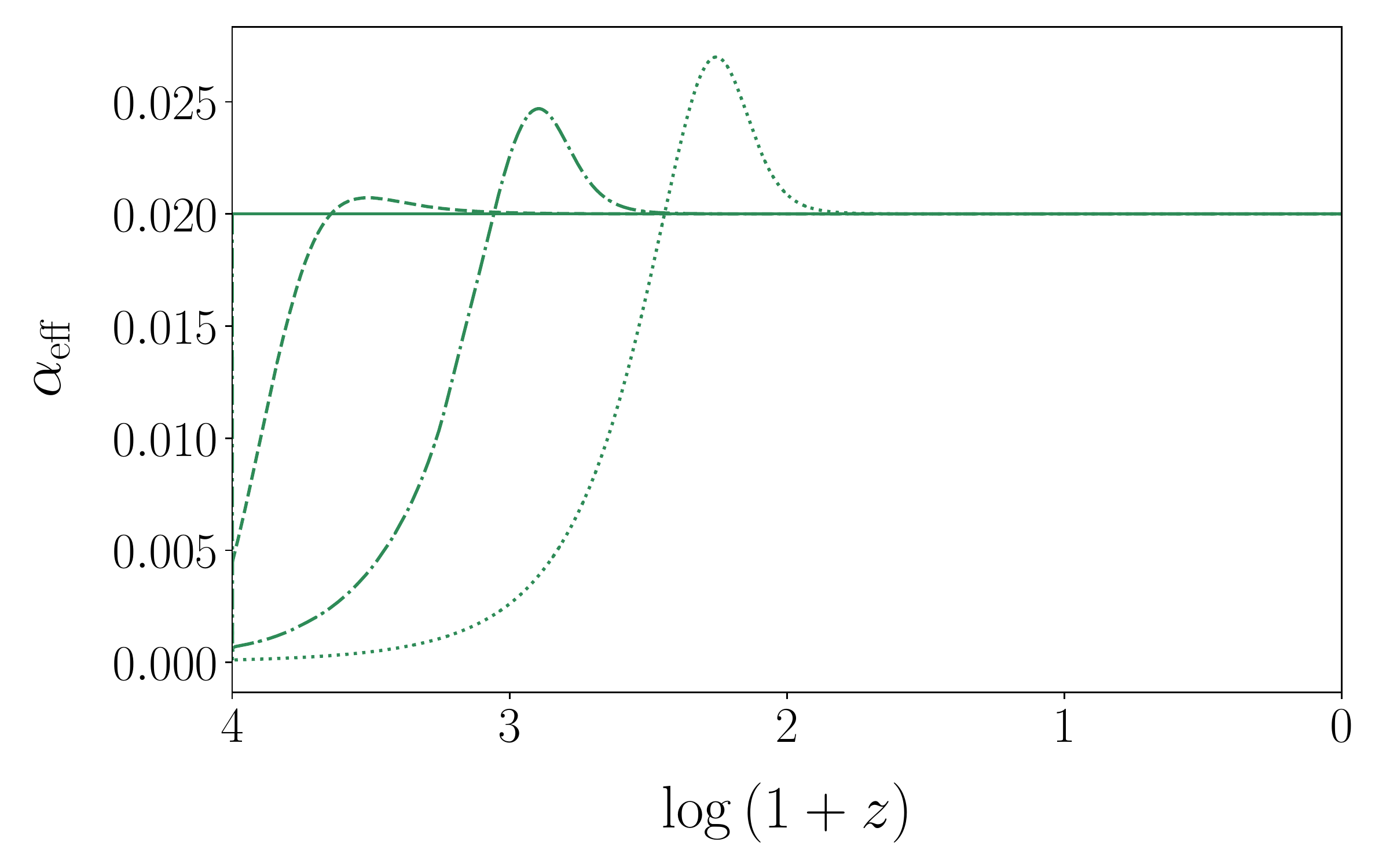}
\caption{\label{fig:sigma2} Left panel, the evolution of $\sigma$ for the models in Table \ref{tabela2}. Model (a), solid line; model (b), dashed line; model (c) dot-dashed line; model (d), dotted line. Right panel, the evolution of the effective conformal coupling $\alpha_{\rm eff}$ for the same models.}
\end{figure*}

The effect of the disformal coupling becomes evident from Fig~\ref{fig:densities2}. As the mass scale $M$ decreases, the disformal contribution becomes more relevant delaying the approach to the conformal kinetic fixed point (B)$_{\rm d}$. Once again, even though $\sigma$ is large at high redshifts, it decays rapidly bringing the disformal coupling to become irrelevant at late times. This is shown in the left panel of Fig.~\ref{fig:sigma2}.
The effective conformal coupling is now depicted in the right panel, where it is evident that decreasing the mass scale, or making the disformal coupling more important, suppresses $\alpha_{\rm eff}$, bringing it close to zero at early times, a result that had already been pointed out in \cite{vandeBruck:2015ida}.

The detailed dynamical analysis performed in the previous sections shows that the kinetic dependence on the disformal function introduces new disformal scaling fixed points that have only a potential transient effect on the early-time cosmology. That is emphasized by this numerical study that reports solely on viable cosmologies that have suppressed late-time disformal contributions, corroborating what has already been discussed in previous studies of the field-dependent disformally coupled model.

\section{Conclusions}

In this work, we have introduced and discussed the construction of a cosmological model, where the role of dark energy is played by a canonical scalar field. Moreover, the dark energy field is allowed to couple to the matter sector by means of a disformal transformation of the metric tensor. We have adopted a dynamical system methodology. More specifically, we have rewritten the main cosmological equations as a dynamical system, constructed from a set of useful (with physical meaning) dimensionless variables. Furthermore, we have identified the invariant sets, including, if necessary, at infinity, and performed a local analysis of the fixed points. Finally, by restricting the parameter space with physical interest, we were able to extract information about the past/future evolution of the Universe.

We have extended the analysis in the existing literature by introducing a dependence on the kinetic term of the scalar field in the disformal factor and performing an extensive and complete dynamical analysis. Since for disformally coupled models, relativistic fluids are also allowed to couple to dark energy, we have started by studying couplings to a single fluid. This allows us to have a first grasp over the past and future asymptotic behaviours associated to this setting. We have assumed a power-law dependence on the kinetic term, expressed through a new parameter, $\mu$. We recall that new disformal solutions arise and, moreover, $\mu$ can be used to shape the parameter region of stability for each fixed point. Depending on the value of the parameters, any of the critical points can have a relevant cosmological role throughout the history of the Universe. Furthermore, we find that there are only two possible late-time attractors capable of reproducing the accelerated expansion of the Universe: a scalar field dominated fixed point and a scaling solution, which again, could be used to alleviate the cosmic coincidence problem. The disformal fixed points can never portray late time acceleration scenarios but, nonetheless, they may be present in the phase space with a transient scaling character. 
%We have studied one example in which the presence of the disformal fixed points translates into a transitional scaling moment that could also address the cosmic coincidence problem. 
We also found candidates for radiation past attractors that are consistent with other coupled quintessence works.
The dynamical analysis was of paramount importance as it provided us with a good understanding of the cosmological features of the model and made it possible to address some of the pathological behaviours found in \cite{Sakstein:2014aca, nelson} (and references therein).

The numerical cosmological analysis of the model leads to the conclusion that, while new early disformal features may be present, the disformal effects are highly inhibited at late times. However, the presence of the disformal coupling, even if small, is valuable to broaden the stability region of the attractors.
It is to be studied in future work, the impact on the evolution of perturbations when $\mu \neq 0$.

%kinetic disformally coupled models have a wide range of phenomenological interest, specially during the matter-dark energy domination transition. Moreover, we have seen that most scalar tensor theories can be related to General Relativity with the addition of a disformally coupled matter sector and can, therefore, be studied and described under this formalism.

\acknowledgments 
NJN and EMT  acknowledge the financial support by  Funda\c{c}\~ao para a Ci\^encia e a Tecnologia (FCT) through the research grants: UID/FIS/04434/2019; PTDC/FIS-OUT/29048/2017 (DarkRipple); COMPETE2020: POCI-01-0145-FEDER-028987 \& FCT: PTDC/FIS-AST/28987/2017 (CosmoESPRESSO) and IF/00852/2015 (Dark Couplings).  EMT was also supported by the grant SFRH/BD/143231/2019 from FCT.

\begin{appendices}

\section{Eigenvalues of the Stability Matrix} \label{app:eigdr}

Here we present the eigenvalues for the dust-like and radiation-like fixed points of the system of equations \eqref{xld}-\eqref{sild} with $\sigma=0$ in Table \ref{table:gamma1vaps} and Table \ref{table:gamma43vaps}, respectively. The naming of the fixed points follows the ones presented in Table \ref{table:gamma1d} for the case of a non-relativistic fluid (labelled as ${d}$) and in Table \ref{table:gamma43m1} for a relativistic fluid (labelled as ${r}$). The eigenvalues of the $3 \times 3$ linear stability matrix $\mathcal{M}$, evaluated at each fixed point $\left( x_f, y_f, \sigma_f \right)$, are used as a tool to infer the stability of the fixed points of the system. The linear stability matrix is constructed from linearisation of the the dynamical system of equations~\eqref{xld}-\eqref{sild}. We choose to not present the eigenvalues for the disformal fixed points ($\sigma \neq 0$) due to their intricate analytical expression. Their stability character is studied semi-analytically, as detailed in the main text (in Sec. \ref{sec:disfdynanal}).

\begin{table*}[ht]
\centering
\begin{tabular}{c c c c}
\hline\hline \\[-1.5ex]
 Name & $e_1$ & $e_2$ & $e_3$ \\ [0.5ex]% inserts table %heading
\hline \\[-1.5ex]
(A$_+$)$_{\rm d}$ &$3 + \sqrt{6} \alpha$ & $3 - \sqrt{\frac{3}{2}} \lambda$ & $2 \left( \sqrt{6} \beta - 3 \left( 1+\mu \right) \right)$ \\
(A$_-$)$_{\rm d}$ & $3- \sqrt{6} \alpha$ & $3 + \sqrt{\frac{3}{2}} \lambda$ & $-2 \left( \sqrt{6} \beta + 3 \left( 1+ \mu \right) \right)$ \\
(B)$_{\rm d}$ &$- \frac{3}{2} + \alpha^2$ & $ \frac{3}{2} + \alpha^2 + \alpha \lambda$ & $-4 \alpha \beta - \left( 3+2 \alpha^2 \right) \left( 1+ \mu \right) $   \\
(C)$_{\rm d}$ & $-3 + \frac{1}{2} \lambda^2$ & $-3 + \lambda \left( \alpha + \lambda \right) $ & $ \lambda \left( 2 \beta - \lambda \left( 1 + \mu \right) \right) $ \\
(D)$_{\rm d}$ & $-\frac{3}{4} \frac{ 2\alpha + \lambda }{ \alpha + \lambda} -\frac{\sqrt{3}}{4} \xi$ & $-\frac{3}{4} \frac{ 2\alpha + \lambda }{ \alpha + \lambda} +\frac{\sqrt{3}}{4} \xi$ & $\frac{3 \left( 2 \beta - \lambda \left( 1 + \mu \right) \right)}{\alpha + \lambda}$\\[1ex]
\hline
\end{tabular}
\caption{Eigenvalues for the fixed points of the system of equations~\eqref{xld}-\eqref{sild} for the $\gamma=1$ case given in terms of the parameters.}
\label{table:gamma1vaps}
\end{table*}

\noindent For the fixed point (D)$_{\rm d}$ in Table \ref{table:gamma1vaps} we have: $\xi= \frac{1}{4 |\alpha+\lambda| } \sqrt{3 \left( 72 + 12 \alpha \left( 5\alpha +3 \lambda \right) - 16 \alpha \lambda \left( \alpha + \lambda \right)^2  - 21 \lambda^2 \right)}.$

\begin{table}[h!]
\centering
\begin{tabular}{c c c c}
\hline\hline \\[-1.5ex]
 Name & $e_1$ & $e_2$ & $e_3$ \\ [0.5ex]% inserts table %heading
\hline \\[-1.5ex]
(O)$_{\rm r}$ & $2$ & $-1$ & $-8$ \\
(A$_\pm$)$_{\rm r}$ & $2$ & $-2 \left(\sqrt{6} \beta +6\right)$ & $\sqrt{\frac{3}{2}} \lambda +3$ \\
(B)$_{\rm r}$ & $-3 + \frac{1}{2} \lambda^2$ & $-4 + \lambda^2 $ & $ 2\lambda \left( \beta - \lambda \right) $ \\
(C)$_{\rm r}$ &$-\frac{1}{2} \left(\sqrt{\frac{64}{ \lambda^2}-15}+1 \right)$ &$\frac{1}{2} \left( \sqrt{\frac{64}{ \lambda ^2}-15}-1 \right)$ & $\frac{8 \beta }{\lambda }-8$ \\[1ex]
\hline
\end{tabular}
\caption{Eigenvalues for the fixed points of the system of equations~\eqref{xld}-\eqref{sild} for the $\gamma=4/3$ and $\mu=1$ case, given in terms of the parameters.}
\label{table:gamma43vaps}
\end{table}

\end{appendices}
%\bibliographystyle{unsrt}

% External bibliography database file in the BibTeX format
\bibliography{mybibD} 

\end{document}